\documentclass[11pt]{JHEP3}
\usepackage{amsmath,amssymb}
\usepackage[T1]{fontenc} % if needed
\usepackage{slashed}
\newcommand{\be}{\begin{equation}}
\newcommand{\ee}{\end{equation}}
\newcommand{\ben}{\begin{eqnarray}}
\newcommand{\een}{\end{eqnarray}}
\newcommand{\nn}{\nonumber}

\newcommand{\ads}{{\text{AdS}_5}}

\newcommand{\NN}{{\cal N}}

\def\one{{\hbox{ 1\kern-.8mm l}}}
\def\zero{{\hbox{ 0\kern-1.5mm 0}}}

\def\t{\tau}
\def\e{\,{\rm e}}

\def\cn{{\mathcal N}}

\def\tr{{\rm Tr}}

\def\l{\left}
\def\r{\right}

\def\s{\sigma}
\def\g{\gamma}
\def\G{\Gamma}
\def\m{\mu}

\def\e{\eta}

\def\o[#1]{{\rm O}\left({#1}\right)}
\def\dotl[#1,#2]{\left\langle #1, #2 \right\rangle}
\def\dotlb[#1,#2]{[ #1, #2 ]}
\def\dotp[#1,#2]{(#1) \cdot (#2)}
\def\>{\rangle}
\def\<{\langle}

\def\ss2{\text{S}^2}
\def\sn{\ss2/\mathbb{Z}_N}
\def\adss{\text{AdS}_2}

\def\s2s2{\ss2\otimes\ss2}
\def\ads2s2{\adss\otimes\ss2}
\def\s2s2n{\l(\ss2\otimes\ss2\r)/\mathbb{Z}_N}
\def\ads2s2n{\l(\adss\otimes\ss2\r)/\mathbb{Z}_N}
\def\zn{\mathbb{Z}_N}

\def\bs{\bar{s}}
\def\ca{\mathcal{A}}

\def\tk{\tilde{k}}
\def\tm{\tilde{m}}

\title{Logarithmic Corrections to Extremal Black Hole Entropy in $\cn=2,4$ and $8$ Supergravity}

\author{Rajesh Kumar Gupta$^a$\footnote{ rgupta AT ictp DOT it}, Shailesh Lal$^b$\footnote{shailesh DOT hri AT gmail DOT com}, Somyadip Thakur$^c$\footnote{somyadip AT cts DOT iisc DOT ernet DOT in}\\

$^a$ $\,$ICTP, High Energy, Cosmology and Astroparticle Physics,\\
$\;$ $\,$Strada Costiera 11, 34151, Trieste, Italy.\\

$^b$ $\,$Department of Physics and Astronomy, \\
$\;$ $\,$Seoul National University \\
$\;$ $\,$Seoul 151-747, Korea \\

$^c$ $\,$Centre for High Energy Physics,\\
$\;$ $\,$Indian Institute of Science, C.V. Raman Avenue,\\
$\;$ $\,$Bangalore 560012, India.\\
}

\abstract{We compute the logarithmic correction to black hole entropy about exponentially suppressed saddle points of the Quantum Entropy Function corresponding to $\zn$ orbifolds of the near horizon geometry of the extremal black hole under study. By carefully accounting for zero mode contributions we show that the logarithmic contributions for quarter--BPS black holes in $\cn=4$ supergravity and one--eighth BPS black holes in $\cn=8$ supergravity perfectly match with the prediction from the microstate counting. We also find that the logarithmic contribution for half--BPS black holes in $\cn=2$ supergravity depends non-trivially on the $\zn$ orbifold. Our analysis draws heavily on the results we had previously obtained for heat kernel coefficients on $\zn$ orbifolds of spheres and hyperboloids in arXiv:1311.6286 and we also propose a generalization of the Plancherel formula to $\zn$ orbifolds of hyperboloids to an expression involving the Harish-Chandra character of $sl\l(2,R\r)$, a result which is of possible mathematical interest.}

\preprint{SNUTP14-002}
\begin{document} 

\section{Introduction}
Since the work of Bekenstein and Hawking, it has been known that black holes have entropy in a quantum theory of gravity. In particular, the entropy of a black hole is determined in terms of the area $A_H$ of its event horizon
\begin{equation}
S_{BH}= {A_H\over 4 G_N},
\end{equation}
where we have set all fundamental constants except the Newton's constant to one. However, it is expected that this formula will receive corrections in a complete theory of quantum gravity from two sources. Firstly, there may be classical corrections from higher derivative terms and secondly, there would be further quantum corrections to this formula, given that it has been obtained in the semi-classical approximation. Incorporation of higher-derivative corrections is achieved by means of the Wald formula \cite{Wald:1993nt}, which is technically challenging but conceptually well-understood. The incorporation of quantum corrections is a more formidable task. In this situation, it is advisable to seek simpler settings in which the problem may potentially be solved as explicitly as possible, with the expectation that the lessons thus learnt would carry over to the more general cases. 

Extremal black holes are an ideal setting to carry out this program. For one, since the near--horizon geometry of an extremal black hole always contains an $\adss$ factor \cite{Kunduri:2007vf,Figueras:2008qh}, their classical entropy formula already admits important simplifications, and can be computed by solving an algebraic set of equations \cite{Sen:2005wa,Sen:2005iz}, a remarkable simplification over the general procedure for computing the Wald entropy. Remarkably, an expression for the full quantum entropy of these black holes has been proposed using the AdS/CFT correspondence, again exploiting the presence of the $\adss$ factor in the near--horizon geometry. In particular, the full quantum degeneracy associated with the event horizon for extremal black holes carrying charges $\vec{q}\equiv q_i$ is proposed to be given by the string path integral \cite{Sen:2008vm,Sen:2008yk}
\begin{equation}\label{qef}
d_{hor}\l(\vec{q}\r)\equiv\l\langle\exp\l[i\oint q_i d\theta\mathcal{A}_\theta^i\r] \r\rangle_{\adss}^{finite},
\end{equation}
where the subscript `finite' means that the volume divergence of the path integral due to the $\adss$ factor has been regulated in accordance with the general principles of the AdS/CFT correspondence. We refer the reader to \cite{Sen:2009vz,Sen:2009md,Sen:2010ts,Banerjee:2008ky,Banerjee:2009uk,Murthy:2009dq,Sen:2009gy} for additional work on this proposal and the lectures \cite{Mandal:2010cj} for a review.

The path integral receives contributions from all field configurations which asymptote to the near horizon geometry of the extremal black hole as is usual in AdS/CFT. This will be reviewed below in more detail. Since the work of Strominger and Vafa \cite{Strominger:1996sh}, in many cases the exact formula for entropy of supersymmetric extremal black holes has been computed in $\cn=4$ and $\cn=8$ string theory, see for example \cite{Dijkgraaf:1996it,Maldacena:1999bp,LopesCardoso:2004xf,Shih:2005uc,Shih:2005qf,Jatkar:2005bh, Dabholkar:2006xa,David:2006ji,David:2006ru,David:2006yn,David:2006ud,Sen:2008ta,Sen:2009gy}. One can therefore try to explicitly check the validity of the proposal of \cite{Sen:2008vm,Sen:2008yk} in this context. More ambitiously, one can try to use \eqref{qef} to compute the quantum entropy of black holes where the microscopic string answer is so far not available. This would be a non-trivial prediction arising from this proposal. Given this motivation, one would be interested in evaluating the path integral \eqref{qef} either exactly using localisation \cite{Banerjee:2009af,Dabholkar:2010uh,Dabholkar:2011ec,Gupta:2012cy}, or perturbatively in a loop expansion \cite{Banerjee:2010qc,Banerjee:2011jp,Gupta:2013sva}. In this paper, we shall adopt the second approach.

In the semiclassical approximation, the dominant contribution to the path integral \eqref{qef} comes from its saddle points. The leading saddle point corresponds to the attractor geometry itself, and it has been shown that the on-shell action evaluated on this saddle point correctly reproduces the Wald entropy of the black hole \cite{Sen:2008vm}. Further, loop corrections about this saddle point have also been computed for $\cn=4$ and $\cn=8$ supergravity to find that \cite{Banerjee:2010qc,Banerjee:2011jp}
\begin{equation}
\begin{split}
S_{BH}&=\frac{A_H}{4 G_N}+{\mathcal O}(1),\quad  {\mathcal N}=4,\\
S_{BH}&=\frac{A_H}{4 G_N}-4\ln{A_H\over G_N}+{\mathcal O}(1), \quad{\mathcal N}=8.
\end{split}
\end{equation}
These expressions precisely match with the prediction from the microscopic answer from string theory \cite{LopesCardoso:2004xf,Jatkar:2005bh,David:2006ud,Sen:2009gy} and therefore provide a non-trivial quantum test of the validity of the quantum entropy function. Similar computations have also been carried out for $\cn=2$ BPS black holes to find \cite{Sen:2011ba}
\begin{equation}
S_{BH}=\frac{A_H}{4G_N}+\l(2-{\chi\over 24}\r)\ln{ A_H\over G_N}+{\mathcal O}(1),
\end{equation}
where $\chi=2\l(n_V-n_H+1\r)$ is the Euler characterestic of the Calabi-Yau three-fold on which IIA string theory is compactified to obtain the $\cn=2$ string theory, and $n_H$ and $n_V$ are the number of hypermultiplets and vector multiplets present in the theory. In this case, the corresponding microscopic result is not known, though \cite{Sen:2011ba} observed an intriguing match with a version of the OSV formula computed in an \textit{a priori} different scaling limit of black hole charges \cite{Denef:2007vg}.

In this paper, we shall study loop corrections about saddle points of \eqref{qef} which correspond to orbifolds of the near horizon geometry, where the quotient group is a $\zn$ subgroup of the $\adss\otimes\ss2$ isometry group $SL(2;{\mathbb R})\times SU(2)$. We shall define the action of the quotient group on the attractor geometry in Section \ref{sec2}. In the microscopic picture, these saddle points are expected to correspond to exponentially suppressed corrections to the asymptotic formula for the statistical degeneracy $d_{micro}$ \cite{Banerjee:2008ky,Banerjee:2009af,Sen:2009gy}. These corrections are certainly present in the microscopic formulae in $\cn=4$ and $\cn=8$ string theory \cite{Banerjee:2008ky,Sen:2009gy}, and the leading contribution to $d_{micro}$ has been computed from both the microscopic formula and the Quantum Entropy Function and matched with each other \cite{Banerjee:2009af}. We shall focus on the next-to-leading contribution, the `log term', in this paper. It is expected on general grounds \cite{Sen:2012dw} that this contribution arises only from one-loop contributions of massless fields, and only from the two-derivative sector of the theory. Therefore, though the path integral \eqref{qef} is over all string fields, this particular contribution may be computed entirely from massless supergravity fields in the two-derivative approximation. We will find that the log term matches perfectly with the microscopic results from string theory in all cases where it is available.

Since the contribution to log terms appears only from the one-loop partition function it takes the form of certain determinants of operators of Laplace type, which may be computed efficiently using the heat kernel method \cite{Vassilevich:2003xt}. This, however, requires us to compute the heat kernel of the kinetic operator corresponding to the graviphoton background over the $\zn$ orbifold of $\adss\otimes\ss2$. This was initiated in \cite{Gupta:2013sva} where an expression for the heat kernel of the Laplacian for vectors, scalars and spin--${1\over 2}$ fields was obtained on these quotient spaces. Then, the heat kernel for a single $\cn=4$ vector multiplet was computed in the graviphoton background, and the log term was shown to vanish, in accordance with the expectation from the microscopic counting. This also required a careful counting of zero modes of the gauge field in the $\cn=4$ vector multiplet. 

In this paper, we shall extend the analysis of \cite{Gupta:2013sva} to include gravitons and spin--${3\over 2}$ fields as well. This will enable us to compute the contribution of the gravity multiplet of $\cn=4$ and $\cn=8$ supergravity, as well as the gravity, vector and hypermultiplets of $\cn=2$ theories. As the computation is involved, we would like to present here a brief overview of the strategy and results obtained. Firstly, since the background we're computing the determinants on is a quotient of the attractor geometry, the analysis of \cite{Banerjee:2010qc} and especially \cite{Banerjee:2011jp} will be of great utility. It was found there that incorporating the graviphoton flux into the heat kernel changes the eigenvalues of the kinetic operator from those of the Laplacian, but the degeneracies do not change\footnote{The notion of `degeneracy' is a little tricky for $\adss$, as any eigenvalue has a countably infinite number of independent eigenfunctions associated with it. However, when the volume divergence of $\adss$ is regulated as described below, `degeneracy' becomes well-defined. It is essentially the Plancherel measure. See \cite{Camporesi:1994ga,Camporesi:1995fb,Camporesi} for more details.} The shift in eigenvalues could be systematically taken into account either exactly, or to the order relevant to extracting the log term. The analysis of this paper is exactly along the lines of \cite{Banerjee:2011jp}. In particular, the kinetic operator and the one-loop determinant being evaluated is exactly the same. The only modification is that we evaluate the determinant the subspace of fields which survive the $\zn$ orbifolding. The effect of the orbifolding is to change the degeneracies of the eigenvalues of the kinetic operator from those on the unquotiented space $\adss\otimes\ss2$ to new ones on $\ads2s2n$. Again, using the same regularisation for the $\adss$ volume, the degeneracy can be made well-defined. We shall use the results of \cite{Gupta:2013sva} to propose and motivate a very simple form for the degeneracy on these quotient spaces. In particular, the group theoretic form for the heat kernel will be especially useful for this purpose. 

This leads to an important simplification in the computation. It was shown in \cite{Gupta:2013sva} that the heat kernel on the orbifold background is a sum of $\frac{1}{N}$ times the heat kernel on the unorbifolded space plus contributions which come from the fixed points of the orbifold. Due to technical reasons which we shall outline, the contribution of the fixed points to logarithmic corrections is independent of the eigenvalues of the kinetic operator and depends only on the degeneracy of the eigenvalue. Hence, this contribution may be directly read off from the heat kernel of the Laplacian itself. An important distinction between the partition function on the unquotiented and the orbifold backgrounds is that of zero modes. Unlike the case of vector zero modes which were dealt with in \cite{Gupta:2013sva}, we find that the number of zero modes of the kinetic operator evaluated on gravitinos and gravitini changes non-trivially. On carefully analysing the contribution of these zero modes to the partition function, one obtains a match with the microscopic results for $\cn=4$ and $\cn=8$ string theory. We finally find the following results for the logarithmic corrections about the orbifold background $AdS_2\times S^2/{\mathbb Z}_N$, 
\begin{equation}
\begin{split}
\ln\l(d_{hor,N}\r)&=\frac{A_H}{4N G_N}+{\mathcal O}(1), \quad {\mathcal N}=4,\\
\ln\l(d_{hor,N}\r)&=\frac{A_H}{4N G_N}-4\ln{A_H\over G_N}+{\mathcal O}(1), \quad{\mathcal N}=8,
\end{split}
\end{equation}
where $d_{hor,N}$ is the contribution to horizon degeneracy as computed via the path integral \eqref{qef} evaluated about the saddle point corresponding to the $\zn$ orbifold of the attractor geometry of the black hole. These results are in perfect agreement with the microscopic computation \cite{Banerjee:2008ky,Sen:2009gy}. Additionally, the analysis may be extended to include half--BPS black holes in $\cn=2$ supergravity, where we find
\begin{equation}
\ln\l(d_{hor,N}\r)=\frac{A_H}{4N G_N}+\l(2-\frac{N\,\chi}{24}\r)\ln{A_H\over G_N}+{\mathcal O}(1).
\end{equation}
This answer has a slightly unexpected feature. In contrast to the results for $\cn=4$ and $\cn=8$ supergravity, the logarithmic correction depends on the orbifold under consideration, and surprisingly the parameter $N$ appears in the numerator. However, it does reduce to the unquotiented answer on setting $N=1$, which on the face of it seems non-trivial as the number of zero modes changes when the orbifold is imposed. These results are summarised in Table \ref{t1}.

We close this section with a brief overview of the rest of the paper. Section \ref{sec2} contains a review of the saddle points of the Quantum Entropy Function that we compute the log term about. Section \ref{sec3} is a review of the heat kernel method as it applies to extraction of the log term. Section \ref{sec4} reviews the group theoretic form for the heat kernel on $\ads2s2n$ and contains a proposed expression for the degeneracy of eigenvalues of the Laplacian on $\ads2s2n$, a result of possible mathematical interest. Sections \ref{sec5}, \ref{sec6}, \ref{sec7} put these results together, along with the zero mode contributions to compute log corrections in $\cn=4$, $\cn=8$, and $\cn=2$ supergravity results, which have been tabulated above in Table \ref{t1}. We then conclude. The Appendices contain some results useful in our analysis. 
\begin{table} { 
\begin{center}\def\st{\vrule height 3ex width 0ex}
\begin{tabular}{|c|c|c|c|c|c|c|c|c|c|c|} \hline 
Theory  & Macroscopic ($\zn$) & Macroscopic (unquotiented) & Microscopic  
\st\\[1ex] \hline \hline
$\NN=4$ & 0  & $0$ & 0 \st\\[1ex] \hline
$\NN=8$ &   $-4$ & $-4$ & $-4$ \st\\[1ex] \hline
\hbox{$\cn=2$} &   $\l(2-\frac{N\,\chi}{24}\r)$ & $\l(2-\frac{\chi}{24}\r)$ & $?$ 
\st\\[1ex] \hline \hline 
\end{tabular}
\caption{The coefficient of the $\ln\l({A_H\over G_N}\r)$ term in $\ln\l(d_{hor,N}\r)$ computed for $\cn=2,4$ and 8 supergravities, both for the $\zn$ orbifold and for the unorbifolded saddle-point, obtained in \cite{Banerjee:2011jp}. We see that for $N=1$ all the $\zn$ results reduce to the unorbifolded case. The microscopic results for the exponentially suppressed saddle point, where available, are also listed for comparison. $\chi$ is the Euler characterestic of the Calabi-Yau three-fold on which type IIA string theory is compactified to obtain the specific $\cn=2$ string theory under study.
} \label{t1}
\end{center} }
\end{table}
\section{Exponentially Suppressed Contributions to Statistical Entropy and Saddle Points of the Quantum Entropy Function}\label{sec2}
This section is a review of asymptotic formulae for black hole entropy as obtained by microstate counting in string theory, along with the proposal of \cite{Sen:2008vm,Sen:2008yk} for the macroscopic origin of this formula. For definiteness, we focus on microscropic results in $\cn=8$ string theory \cite{Sen:2009gy}, the $\cn=4$ case has already been reviewed in \cite{Gupta:2013sva}. As our focus is just on the overall structure of the asymptotic expansion for microscopic degeneracy, we shall be heuristic here and refer the reader to the original papers \cite{Banerjee:2008ky,Sen:2009gy} for details and explanations. We shall consider, following \cite{Sen:2009gy}, the $\cn=8$ string theory obtained by compactifying Type IIB string theory on $T^6$. In this case the asymptotic formula for degeneracy of a dyon carrying charges $Q$ and $P$ takes the form \cite{Sen:2009gy}
\begin{equation}\label{asymptoticmicro}
d_{micro}(Q,P)\simeq\l(-1\r)^{Q\cdot P+1}\sum_{s} s \l(-1\r)^{{\Delta\over s^2}+1}\l({\Delta\over s^2}\r)^{-2}e^{{\pi\sqrt{\Delta}}\over s}\equiv \sum_s d_{micro,s}\l(Q,P\r),
\end{equation}
where $s$ is an integer which obeys certain conditions explicitly enumerated in \cite{Sen:2009gy}, and the Cremmer-Julia invariant, $\Delta$, is given by \cite{cremmerjulia,kalloshkol}
\begin{equation}
\Delta\l(Q,P\r)=Q^2P^2-\l(Q\cdot P\r)^2.
\end{equation}
The leading contribution to \eqref{asymptoticmicro} comes from the $s=1$ term in the sum. This corresponds to the usual Bekenstein-Hawking entropy on the macroscopic side. The terms with higher values of $s$ correspond to exponentially suppressed corrections to this leading answer. We will now review the proposal for the origin of these exponentially suppressed corrections within the framework of the quantum entropy function.

The quantum entropy function \eqref{qef} is defined as the string path integral, with a Wilson line insertion, in asymptotically $\adss$ spacetimes. We now briefly review some saddle points of this path integral. We refer the reader to \cite{Banerjee:2009af} for more details and a systematic classification of these saddle points. We consider, following \cite{Sen:2009gy}, Type IIB string theory compactified on $T^4$ and six dimensional geometries that are asymptotically $\adss\otimes\ss2\otimes S_1\otimes\tilde{S}_1$. The simplest saddle point of the string path integral is the attractor geometry itself, which is $\adss\otimes\ss2\otimes S_1\otimes\tilde{S}_1$ supported by fluxes.
\begin{equation}\label{attractor}
\begin{split}
ds^2 &= a\l(d\eta^2+\sinh^2\eta d\theta^2\r) + a\l(d\psi^2+\sin^2\psi d\phi^2\r) + {R^2\over\tau_2} \vert dx^4 +\tau dx^5\vert^2,\\
G^I  &= {1\over 8\pi^2}\l[Q_I \sin\psi dx^5\wedge d\psi\wedge d\phi + P_I \sin\psi dx^{4}\wedge d\psi\wedge d\phi + \text{dual}\r],\\
V_I^i &=\text{constant},\quad V_I^r = \text{constant}.
\end{split}
\end{equation}
Here $a$ is the $\adss$ and $\ss2$ radius, and $R$ is a real number which does not scale with $a$. $\t$ is a complex number which also does not scale with $a$, $V$ is a matrix--valued scalar field. We refer the reader to \cite{Sen:2009gy} for more details regarding this geometry. The construction of the saddle point in $\cn=4$ supergravity is entirely analogous, the only difference is that IIB string theory is now compactified on K3 instead of $T^4$ \cite{Banerjee:2009af}.

Now it may be shown that the leading contribution to the quantum entropy function from this saddle point is precisely $e^{S_{Wald}}$ where $S_{Wald}$ is the Wald entropy of the extremal black hole. Further, by expanding in fluctuations about this saddle point, the contribution to the quantum entropy function which scales as log($A_H$) may be matched from the microscopic result \cite{Banerjee:2010qc, Banerjee:2011jp}. Our focus in this paper will be field configurations which correspond to $\zn$ orbifolds of the attractor geometry \eqref{attractor}, where the quotient group acts via
\begin{equation}\label{orbact}
\l(\theta,\phi,x^5\r)\mapsto\l(\theta+{2\pi\over N},\phi-{2\pi\over N},x^5+{2\pi k\over N}\r),\quad k,N\in\mathbb{Z},\quad \text{gcd}\l(k,N\r)=1.
\end{equation}
This is a freely acting orbifold of \eqref{attractor}. It is possible to show by a change of coordinates that this orbifold obeys the same asymptotic boundary conditions as \eqref{attractor}. It is therefore an admissible saddle point of the quantum entropy function. Additionally, the classical contribution to the quantum entropy function is $e^{S_{Wald}\over N}$. Further, though these orbifolds break supersymmetry partly, the integration over the zero modes corresponding to broken symmetries does not make the path integral vanish automatically \cite{Banerjee:2009af}. Therefore, this orbifold is a promising candidate to correspond to the exponentially suppressed terms in \eqref{asymptoticmicro} if $s$ and $N$ are identified to each other.

In this paper we shall perform a further test of this conjecture along the lines of \cite{Banerjee:2010qc} and \cite{Banerjee:2011jp}. In particular, note that for any value of $s$,
\begin{equation}
\ln\l(d_{micro,s}\r)\simeq {\pi\sqrt{\Delta}\over s} -2\ln\Delta +\mathcal{O}\l(1\r).
\end{equation}
In particular, we shall reproduce the second term in this expansion from quadratic fluctuations about the quotient \eqref{orbact} of \eqref{attractor}. We carry out first the same computation in $\cn=4$ supergravity, where the term proportional to $\ln\Delta$ is known to vanish from the microscopic side, and then extend our computation to $\cn=8$ supergravity. We find that our macroscopic results match perfectly with the predications from the microscopic theory. We then carry out the corresponding computation for $\cn=2$ supergravity.

\section{The Heat Kernel Method}\label{sec3}
We have reviewed how the saddle points of the Quantum Entropy Function corresponding to $\zn$ orbifolds of the black hole attractor geometry are promising candidates to reproduce the exponentially suppressed contributions to the statistical dyon degeneracy computed from $\cn=4$ and $\cn=8$ string theory in the large charge limit. In particular, they correctly reproduce the leading growth of the exponentially suppressed saddle points of the microscopic degeneracy formula. The goal of this paper is to show that the next to leading growth in charges is also correctly reproduced by these orbifolds. On the microscopic side, the subleading term corresponds to a contribution to $\ln\l(d_{micro,N}\r)$ which goes as $\ln\Delta$. To compare this from the macroscopic side, we need to extract the $\ln\l(a\r)$ term from the contribution of the saddle point \eqref{attractor}, \eqref{orbact} to the partition function \eqref{qef}. For this purpose, we will think of the string theory that \eqref{attractor} is embedded in as a supergravity theory on $\adss\otimes\ss2$ coupled to an infinite number of massive fields which may be stringy or Kaluza-Klein modes. This leads to some remarkable simplifications on very general grounds \cite{Sen:2012dw}. In particular, it  may be shown that the term which scales as $\ln\l(a\r)$ in the partition function can arise only from one--loop fluctuations of massless fields in the four-dimensional geometry \cite{Banerjee:2010qc,Banerjee:2011jp,Sen:2012dw}. The heat kernel method is therefore ideally suited to analyse this problem, and we review it briefly here in order to set up notation and remind the reader of a few salient facts, mostly following the discussion in \cite{Bhattacharyya:2012ye}. We refer the reader to \cite{Vassilevich:2003xt} for a general overview of the heat kernel method, and to \cite{Bhattacharyya:2012ye,Gupta:2013sva} for more details regarding the discussion below. The one-loop partition function for a $d+1$ dimensional theory with overall length scale $a$ takes the form
\begin{equation}
\mathcal{Z}_{1-\ell}= \l({\det}'\l(M\r)\r)^{-{1\over 2}}\cdot\mathcal{Z}_{zero}\l(a\r),
\end{equation}
where $M$ is an operator which in our case is a positive semi-definite operator whose eigenvalues scale as ${1\over a^2}$. The prime indicates that the determinant is evaluated on fields which are not zero modes of the operator $M$, $\mathcal{Z}_{zero}$ is the zero mode contribution to the partition function, and the argument $a$ in $\mathcal{Z}_{zero}$ reminds us that the zero mode contribution also scales non-trivially with $a$. Now the determinant of $M$ may be related to its trace via
\begin{equation}
\ln{\det}'\l(M\r) = \,-\int_0^\infty {dt\over t}\, \tr'\l(e^{-t M}\r).
\end{equation}
Now the trace can be conveniently computed by means of the integrated heat kernel
\begin{equation}\label{kernel}
K(t) = \tr\l(e^{-t M}\r) = \sum_{n}\sum_{m=1}^{\tilde{m}_n} \int_{\mathcal{M}}d^{d+1}x\sqrt{g} \psi^*_{n,m}\l(x\r)\psi_{n,m} e^{-{t\over a^2} E_n},
\end{equation}
where $\psi_{n,m}$ are eigenfunctions of the operator $M$ belonging to the $\tm_n$-fold degenerate eigenvalue $E_n$. Note that this expression is perfectly well-defined for compact manifolds like $\ss2$, but is naively divergent for non-compact spaces like hyperboloids. However, it is by now well-understood how this divergence may be regulated in accordance with the general principles of the AdS/CFT correspondence and a sensible answer extracted for $K(t)$ \cite{Banerjee:2010qc,Banerjee:2011jp,Sen:2012cj}. This involves putting a cutoff on the radial coordinate of global $\adss$ and extracting the term which is of order 1 in the large $\eta_0$ expansion. This procedure also extends nicely to quotients of hyperboloids both with and without fixed points \cite{Gopakumar:2011qs, Gupta:2013sva}. We also note that the heat kernel $K(t)$ in \eqref{kernel} is evaluated over all the eigenfunctions of $M$, including zero modes. To obtain the determinant over non-zero modes, one has to subtract out the zero mode contribution
\begin{equation}
\ln{\det}'\l(M\r) = -\int_0^\infty {dt\over t}\l(K\l(t\r)-n^0_M\r),
\end{equation}
where $n^0_M$ is the number of zero modes $\psi_{0,m}$ of the operator $M$, defined via
\begin{equation}\label{nzero}
n^0_M = \sum_{m=1}^{n^0_M} \int_{\mathcal{M}}d^{d+1}x\sqrt{g} \psi^*_{0,m}\l(x\r)\psi_{0,m},\quad M\vert\psi_{0,m}\rangle=0.
\end{equation}
Again, this procedure is well-defined for compact manifolds and, using the regularisation procedure mentioned above, can be made well-defined for AdS and its quotients as well. Now the heat kernel has a small $t$ expansion of the form 
\begin{equation}
K\l(t\r)={1\over \l(4\pi\r)^{d+1\over 2}}\sum_{n=0}^\infty t^{n-{d+1\over 2}}\int_{\mathcal{M}} d^{d+1}x\sqrt{g}\, a_n(x),
\end{equation}
where a few leading coefficients $a_n$ are known explicitly for the Laplacian on both smooth and conical spaces \cite{Vassilevich:2003xt, Fursaev:1996uz}. Now we may extract the contribution from $\ln{\det}'M$ that scales as $\ln a$. It turns out that the term that scales as $\ln a$ in this determinant receives contributions only from the $t^0$ term in the heat kernel expansion. This will be important in our subsequent analysis. That is
\begin{equation}\label{lndetprime}
\ln{\det}'M = \l({1\over \l(4\pi\r)^{d+1\over 2}}\int_{\mathcal{M}} d^{d+1}x\sqrt{g}\, a_{d+1\over 2}(x)-n^0_M\r)\ln a +\ldots,
\end{equation}
where we have been careful to subtract out the contribution of the zero mode from the heat kernel. The dots denote extra terms which do not scale as $\ln a$. 

We conclude the section with a discussion of the contribution of zero modes to the $\ln(a)$ term to $\ln\l(\mathcal{Z}_{1-\ell}\r)$. This has been analysed on very general grounds in \cite{Sen:2012dw} and we quote the final result. The zero mode contribution from a field $\phi$ to the path integral can be shown to scale as \cite{Banerjee:2010qc,Banerjee:2011jp,Sen:2012dw}
\begin{equation}\label{lnzero}
\mathcal{Z}_{zero}= \l(a^{n_{\phi}\beta_{\phi}}\r)\mathcal{Z}_0,
\end{equation}
where $\mathcal{Z}_0$ does not scale with $a$, $n_\phi$ is the number of zero modes of the operator $M$ evaluated over the field $\phi$ and $\beta_\phi$ is a number which can be computed for every field on a case--by--case basis. For a vector field $\beta_v=1$ \cite{Banerjee:2010qc}, for spin--${3\over 2}$ fields, $\beta_{3\over 2}=3$ and for the graviton $\beta_g=2$ \cite{Banerjee:2011jp}. In that case, using equations \eqref{lndetprime} and \eqref{lnzero}, we can show that
\begin{equation}
\ln\l(\mathcal{Z}_{1-\ell}\r)=\l(K\l(0;t\r)-n^0_M\r)\ln\l(a\r)+\sum_{\phi}n^0_{\phi}\beta_{\phi}\ln\l(a\r)+\ldots,
\end{equation}
where $K\l(0;t\r)$ is the $t^0$ term in the heat kernel expansion when the heat kernel is evaluated over all the eigenmodes of the operator $M$, including zero modes. Again `$\ldots$' denote terms that do not scale as $\ln\l(a\r)$. Using the fact that $\sum_{\phi}n^0_{\phi}=n^0_M$, the total number of zero modes of $M$, we find that the total contribution to the partition function $Z$ proportional to $\ln\l(a\r)$ when expanded about the given saddle point is 
\begin{equation}\label{logcontribution}
\ln\l(\mathcal{Z}\r)_{log}= \l(K\l(0;t\r) + \sum_{\phi}n^0_{\phi}\l(\beta_{\phi}-1\r)\r)\ln\l(a\r).
\end{equation}
We also note here that the number of zero modes \eqref{nzero} can be read off from the coefficient of the $e^{0}$ term in the full heat kernel expansion \eqref{kernel}.
\section{The Heat Kernel on $\ads2s2n$}\label{sec4}
In this section we will review the results we previously obtained in \cite{Gupta:2013sva} for the scalar and spin$-{1\over 2}$ heat kernels in $\ads2s2n$. It will turn out that this analysis will be enough to extract the log term corresponding to the determinant of the kinetic operator evaluated over the $\zn$ orbifold \eqref{orbact} of the graviphoton background \eqref{attractor}. We will find an interesting simplification in the $t^0$ behaviour of the heat kernel which allows us to relate the heat kernel of the rediagonalised fields on this orbifold of the black hole near horizon geometry to the unorbifolded heat kernel plus some extra `conical' terms which can be very simply computed. We now define the geometry and the orbifold projection that we shall impose. In global coordinates, the metric on $\adss\otimes\ss2$ is given by
\begin{equation}
ds^2=a^2\l(d\eta^2+\sinh\eta^2 d\theta^2\r)+ a^2\l(d\rho^2+\sin^2\rho d\phi^2\r),
\end{equation}
where $\l(\eta,\theta\r)$ are coordinates on $\adss$ and $\l(\rho,\phi\r)$ are coordinates on $\ss2$. We shall compute the heat kernel on the space obtained by imposing the following $\zn$ orbifold on $\adss\otimes\ss2$
\begin{equation}\label{znorb}
\l(\theta,\phi\r)\mapsto \l(\theta + {2\pi\over N},\phi-{2\pi\over N}\r).
\end{equation}
We refer to the space thus obtained as $\ads2s2n$. This space is related by analytic continuation to the $\zn$ orbifold \eqref{znorb} of the space $\ss2\otimes\ss2$
\begin{equation}
ds^2=a_1^2\l(d\chi^2+\sin\chi^2 d\theta^2\r)+ a_2^2\l(d\rho^2+\sin^2\rho d\phi^2\r),
\end{equation}
where we have to continue $\chi\mapsto i\eta$ and $\l(a_1,a_2\r)\mapsto\l(ia,a\r)$ to reach $\ads2s2n$. Our strategy in this paper will be the same as that adopted in \cite{Banerjee:2010qc,Banerjee:2011jp,Gupta:2013sva}. We will express vectors, gravitini and gravitons in a basis of states obtained by acting derivatives and gamma matrices on scalars and Dirac fermions, as applicable. This is outlined partially in Appendices \ref{spin2zerosec} and \ref{discretesec}, and we further refer the reader to \cite{Banerjee:2010qc,Banerjee:2011jp} for complete details. For this reason, we shall now concentrate on the heat kernels over over scalars and Dirac fermions on $\ads2s2n$, reviewing the expressions obtained in \cite{Gupta:2013sva}.
\subsection{The Scalar Heat Kernel}
Our strategy for computing the heat kernel of the Laplacian over scalar fields on $\ads2s2n$ was to evaluate the heat kernel on $\s2s2n$ and analytically continue the result to $\ads2s2n$. This was done both explicitly and by observing certain group theoretic properties of the heat kernel on these quotient spaces and proposing an appropriate analytic continuation. In this section we shall briefly review the latter approach and build slightly on it. We begin with the heat kernel of the scalar on $\s2s2n$, which is given by \cite{Gupta:2013sva}
\begin{equation}\label{scalars2s2}
K^s = {1\over N}K^s_{\ss2\otimes\ss2} + {1\over N}\sum_{m=1}^{N-1}\sum_{\ell,\tilde{\ell}=0}^\infty\chi_{\ell,\tilde{\ell}}\l({\pi m\over N}\r)e^{-\bs_1E_\ell}e^{-\bs_2E_{\tilde{\ell}}},
\end{equation}
where 
\begin{equation}
\chi_{\ell,\tilde{\ell}}\l({\pi m\over N}\r)= {\sin\l({\l(2\ell+1\r)\pi m\over N}\r)\over \sin\l({\pi m\over N}\r)} {\sin\l({\l(2\tilde{\ell}+1\r)\pi m\over N}\r)\over \sin\l({\pi m\over N}\r)}
\end{equation}
is the Weyl character in the $\l(\ell,\tilde{\ell}\r)$ representation of $SU(2)\times SU(2)$, 
$\bs_1={t\over a_1^2}$, $\bs_2={t\over a_2^2}$, and $a_1$ and $a_2$ are the radii of the two spheres. $K^s_{\ss2\otimes\ss2}$ is the heat kernel on the unquotiented space $\ss2\otimes\ss2$. Now a simplification that will be important for the rest of the analysis is the following. The sum over $m$ in \eqref{scalars2s2} is finite as $\bs_{1,2}$ approach zero (i.e. $t$ approaches zero). In particular, the order $t^0$ term from the fixed points is given by
\begin{equation}\label{sinsums}
{1\over N}\sum_{m=1}^\infty\sum_{\ell,\tilde{\ell}=0}^\infty \chi_{\ell,\tilde{\ell}}\l({\pi m\over N}\r)={\l(N^2-1\r)\l(N^2+11\r)\over 180 N}.
\end{equation}
We refer the reader to Appendix \ref{usefulsums} for details. We therefore obtain
\begin{equation}\label{kscals2s2na}
K^s_{\s2s2n}\l(\bs_1,\bs_2\r)= {1\over N}K^s_{\ss2\otimes\ss2}+{\l(N^2-1\r)\l(N^2+11\r)\over 180 N} +\mathcal{O}\l(t\r).
\end{equation}
Importantly, we observe that the fixed point contributions to the heat kernel in \eqref{scalars2s2}, associated with the $m\in [1,N-1]$ terms there, are completely finite in the $t$ goes to zero limit. All the ${1\over t}$ behaviour comes from the global $m=0$ contribution. Therefore, firstly it is justified to do a naive power-series expansion in $t$ for the $m\in [1,N-1]$ terms. Secondly the $t^0$ term in the expansion of these terms, which is the only term that contributes to the log term, is completely independent of the eigenvalues. Hence if $m$ scalars on $\s2s2n$ mix among each other due to a background flux, leading to a shift in the eigenvalues, the contribution of the fixed point terms to the $t^0$ term of the heat kernel is always just
\begin{equation}
K_{conical}^{log}={m\over N}\sum_{s=1}^{N-1}\sum_{\ell,\tilde{\ell}=0}^\infty \chi_{\ell,\tilde{\ell}}\l({\pi s\over N}\r)=m K_{conical,\, 1\, scalar}^{log}.
\end{equation}
All the effects of rediagonalisation of the quadratic operator manifest themselves in the `global' $s=0$ term only. This is the main simplification we will use. 

The second feature of the expression \eqref{scalars2s2} which we would like to highlight is the following. The heat kernel $K(t)$ is essentially the trace of the exponential of the Laplacian. Therefore, on general grounds it should be expressible as
\begin{equation}\label{kerneldeg}
K(t) = \tr\l(e^{-t\Delta}\r) = \sum_n d_n e^{-t E_n},
\end{equation}
where $d_n$ is the degeneracy of the eigenvalue $E_n$, which we define via
\begin{equation}
d_n= \sum_{m}\int_{\mathcal{M}}\psi_{n,m}^*\psi_{n,m},
\end{equation}
where $\int_{\mathcal{M}}$ denotes integration over the spacetime manifold that the fields live on, and $\psi_{n,m}$ span a normalized basis of states in the vector space $\mathcal{V}_{n}$ consisting of eigenstates of the Laplacian belonging to the eigenvalue $E_n$\footnote{This is a completely well defined expression for compact manifolds and once the volume divergence is regularized, gives well-defined and sensible answers for $\adss$ and its quotients as well. For the case of discrete modes on $\ads2s2n$, we have have computed degeneracies via this expression very explicitly for vectors \cite{Gupta:2013sva} and, in this paper, spin$-2$ and spin$-{3\over 2}$ fields.}. Comparing \eqref{kerneldeg} with \eqref{scalars2s2}, we find that the degeneracy of the eigenvalue $E_{\ell,\tilde{\ell}}={1\over a_1^2}\ell\l(\ell+1\r)+{1\over a_2^2}\tilde{\ell}\l(\tilde{\ell}+1\r)$ is given by
\begin{equation}
d_{\ell,\tilde{\ell}}={1\over N}\sum_{\ell,\tilde{\ell}=0}^\infty \l(2\ell+1\r)\l(2\tilde{\ell}+1\r)+{1\over N}\sum_{m=1}^{N-1}\sum_{\ell,\tilde{\ell}=0}^\infty\chi_{\ell,\tilde{\ell}}\l({\pi m\over N}\r).
\end{equation}
We now outline how the expressions \eqref{scalars2s2} and \eqref{kscals2s2na} may be analytically continued to the AdS case. This is again a review of the arguments of \cite{Gupta:2013sva}. Firstly we note that the $m=0$ term in the heat kernel expression is just ${1\over N}$ times the heat kernel on the unquotiented space $\adss\otimes\ss2$, therefore, it can be replaced by the scalar heat kernel $K^s_{\adss\otimes\ss2}$, which was explicitly evaluated in \cite{Banerjee:2010qc} by relating it to the coincident heat kernel. In particular, it was found that \cite{Banerjee:2010qc}
\begin{equation}
K^s_{\adss\otimes\ss2}= {\text{Vol.}_{\adss\otimes\ss2}\over 8\pi^2 a^4}\sum_{\ell=0}^\infty \int_0^\infty d\lambda\,\lambda\tanh\l(\pi\lambda\r)\,\l(2\ell+1\r)e^{-\bs\l(\l(\lambda^2+{1\over 4}\r)+\ell\l(\ell+1\r)\r)}.
\end{equation}
Here the volume of $\adss\otimes\ss2$ is divergent as $\adss$ is non-compact. It may be regulated by putting a cutoff on the $\adss$ radial coordinate $\eta$ at a large value $\eta_0$ and retaining the term which is of order 1 in the large $\eta_0$ expansion. Physically, this is because the term divergent in the large $\eta_0$ limit represents a shift in the ground state energy, and the one-loop correction to black hole entropy is contained in the finite part \cite{Sen:2008vm,Sen:2008yk}. Then the regularised volume of $\adss\otimes\ss2$ is found to be \cite{Banerjee:2010qc} 
\begin{equation}
\text{Vol.}_{\adss\otimes\ss2} = -8\pi^2 a^4,
\end{equation}
and the (regulated) heat kernel becomes
\begin{equation}
K^s_{\adss\otimes\ss2}= -\sum_{\ell=0}^\infty \int_0^\infty d\lambda\,\lambda\tanh\l(\pi\lambda\r)\,\l(2\ell+1\r)e^{-\bs\l(\l(\lambda^2+{1\over 4}\r)+\ell\l(\ell+1\r)\r)}.
\end{equation}
Comparing this to \eqref{kerneldeg} we see that on regulating the volume divergence in the heat kernel, we have obtained a well defined measure of degeneracy of the eigenvalue $E_{\lambda\ell}={1\over a^2}\l(\l(\lambda^2+{1\over 4}\r)+\ell\l(\ell+1\r)\r)$ of the scalar Laplacian on $\adss\otimes\ss2$. It is given by
\begin{equation}
\tilde{d}_{\lambda\ell} = -\lambda\tanh\l(\pi\lambda\r)\,\l(2\ell+1\r),
\end{equation}
which is essentially the Plancherel measure. Now the rest of the terms in \eqref{kscals2s2na} may be continued to $\adss\otimes\ss2$ by continuing the $\ss2$ radii $\l(a_1,a_2\r)$ to $\l(ia,a\r)$ and multiplying by an overall half \cite{Gupta:2013sva}. This is because under the $\zn$ orbifold that we're imposing, $\adss$ has half the number of fixed points as $\ss2$. In that case, \eqref{kscals2s2na} gets analytically continued to \cite{Gupta:2013sva}
\begin{equation}\label{kscalads2s2n}
K^s_{\ads2s2n}\l(\bs\r)= {1\over N}K^s_{\adss\otimes\ss2}+{1\over 2}\cdot{\l(N^2-1\r)\l(N^2+11\r)\over 180 N} +\mathcal{O}\l(\bs\r).
\end{equation}
Additionally, motivated by the analytic continuation carried out in \cite{Gopakumar:2011qs} for thermal AdS and other freely acting orbifold groups, we also proposed that an equivalent way of carrying out the analytic continuation would be to replace the Weyl character corresponding to the $\ss2$ being analytically continued to $\adss$ by the Harish-Chandra (global) character\footnote{This differs from the measure on $\lambda$ which we had originally suggested in \cite{Gupta:2013sva} by ${1\over\sin{\pi s\over N}}$, a $\lambda$--independent quantity. The extra factor is necessary to match \eqref{scalarads2s2} with \eqref{kscalads2s2n} for arbitrary $N$, not just $N=2$.} \cite{Lang}
\begin{equation}
\chi^b_{\lambda}\l({\pi s\over N}\r)= {\cosh\l(\pi-{2\pi s\over N}\r)\lambda\over \cosh\l(\pi\lambda\r)\sin\l({\pi s\over N}\r)},
\end{equation}
and multiplying the conical terms by half. We therefore obtain
\begin{equation}\label{scalarads2s2}
K^s_{\ads2s2n}\l(\bs\r)= {1\over N}K^s_{\adss\otimes\ss2} + {1\over 2 N}\sum_{s=1}^{N-1}\sum_{\ell=0}^\infty\int_0^\infty d\lambda\,\chi^b_{\lambda\,\ell}\l({\pi s\over N}\r)e^{-\bs E_{\lambda\ell}},
\end{equation}
where we further defined
\begin{equation}
\chi^b_{\lambda\,\ell}\l({\pi s\over N}\r)=\chi^b_{\lambda}\l({\pi s\over N}\r)\chi_{\ell}\l({\pi s\over N}\r).
\end{equation}
By regulating the volume divergence in the first term of \eqref{scalarads2s2} as was done for the unquotiented example, we find that a measure of the degeneracy of eigenvalues $E_{\lambda\ell}$ is given by
\begin{equation}\label{scalardeg}
d_{\lambda\ell}=-{1\over N}\lambda\tanh\l(\pi\lambda\r)+{1\over 2 N}\sum_{s=1}^{N-1}\chi^b_{\lambda\,\ell}\l({\pi s\over N}\r).
\end{equation}
Therefore, as in the case of the compact manifold $\ss2\otimes\ss2$, the effect of the $\zn$ orbifold is to change the degeneracy of the eigenvalue $E_{\lambda\ell}$ from $\tilde{d}_{\lambda\ell}$ to $d_{\lambda\ell}$. Both $\tilde{d}_{\lambda\ell}$ and $d_{\lambda\ell}$ are perfectly well-defined once the volume divergence has been regulated as above. 

Also, from the above expressions, it is manifest that the conical terms in the heat kernel \eqref{scalarads2s2} are finite in the $t\mapsto 0$ limit. Their contribution to the $t^0$ term, which is the only term in the heat kernel expansion relevant for computing the log term, in the heat kernel on $\ads2s2n$ is therefore insensitive to the precise form of the eigenvalues $E_{\lambda\ell}$. This is what we will use when we consider the full kinetic operator of the gravity multiplet fields of $\cn=4$ supergravity and the $\cn=8$ fields in the graviphoton background.
\subsection{The Fermion Heat Kernel}
We now review the computation of the heat kernel of the Dirac operator on $\ads2s2n$ carried out in \cite{Gupta:2013sva}. As the analysis is entirely analogous to the scalar, we will mostly enumerate main steps and the final results. The heat kernel for the Dirac operator is computed by evaluating the heat kernel of its square\footnote{This should lead to a factor of half on taking the square root. But when we evaluate the operator on Dirac fermions, there is an additional factor of two as Dirac fermions are expanded in the basis of eigenvectors of $\slashed{D}$ with complex coefficients. These two factors cancel. For Majorana fermions, there is indeed an overall factor of half which we shall have to include.}. We will denote this heat kernel as $K^f$. Now since the square of the eigenvalues of the Dirac operator on $\ss2\otimes\ss2$ or $\adss\otimes\ss2$ is the sum of the squares of the eigenvalues of the Dirac operator on $\ss2$ (or $\adss$) and $\ss2$, we thus obtain
\begin{equation}
K^f_{\adss\otimes\ss2}=-K^f_{\adss}K^f_{\ss2},
\end{equation}
and similarly for $\ss2\otimes\ss2$. The overall minus sign is on account of the Grassmann--odd nature of fermions. $K^f$ was explicitly evaluated on the quotient space $\s2s2n$ in \cite{Gupta:2013sva}. It is given by
\begin{equation}
K^{f}_{\s2s2n}\l(a_1,a_2\r)={1\over N}K^f_{\ss2\otimes\ss2}-{4\over N}\sum_{s=1}^{N-1}\sum_{\ell,\ell'=0}^\infty\chi_{\ell+{1\over 2},\ell'+{1\over 2}}\l({\pi s\over N}\r)e^{-t E_{\ell,\ell'}},
\end{equation}
where $E_{\ell,\ell'}={1\over a_1^2}\l(\ell+1\r)^2+{1\over a_2^2}\l(\ell'+1\r)^2$, and $\chi_{\ell+{1\over 2},\ell'+{1\over 2}}$ is the product of $SU(2)$ characters in the spin-$\ell$ and spin-$\ell'$ representations. As for the scalar, we analytically continue this expression by replacing one of the $SU(2)$ Weyl characters by the Harish-Chandra character $\chi^f_\lambda$ for $sl(2,R)$ and multiplying the conical terms by half. This character is given by
\begin{equation}
\chi^{\,f}_\lambda\l({\pi s\over N}\r)={\sinh\l(\pi-{2\pi s\over N}\r)\lambda\over \sinh\l(\pi\lambda\r)\sin\l({\pi s\over N}\r)}.
\end{equation}
Then the heat kernel for the fermion on $\ads2s2n$ is given by
\begin{equation}
K^{f}_{\ads2s2n}={1\over N}K^f_{\adss\otimes\ss2}-{2\over N}\sum_{s=1}^{N-1}\sum_{\ell=0}^\infty\int_0^\infty d\lambda\,\chi^{\,f}_{\lambda}\l({\pi s\over N}\r)\,\chi_{\ell+{1\over 2}}\l({\pi s\over N}\r)e^{-{t\over a^2}E_{\lambda\ell}}.
\end{equation}
We may further check that the above expression for $K^f$ reduces to
\begin{equation}\label{diracads2s2n}
K^{f}_{\ads2s2n}={1\over N}K^f_{\adss\otimes\ss2}-{2\over N}\sum_{s=1}^{N-1}\sum_{\ell'=0}^\infty\sum_{\ell=0}^\infty\chi_{\ell+{1\over 2},\ell'+{1\over 2}}\l({\pi s\over N}\r) + \mathcal{O}\l(t\r).
\end{equation}
We thus see that the $\mathcal{O}\l(t^0\r)$ contribution from the fixed points is independent of the actual form of the eigenvalues $E_{\lambda\ell}$ as for the scalar case. Again, this leads us to define the following degeneracy of eigenvalues $E_{\lambda\ell}$
\begin{equation}\label{fermiondeg}
d_{\lambda\ell}=-{1\over N}\lambda\coth\l(\pi\lambda\r) + {2\over N}\sum_{s=1}^{N-1} {\sinh\l(\pi-{2\pi s\over N}\r)\lambda\over \sinh\l(\pi\lambda\r)\sin\l({\pi s\over N}\r)}\chi_{\ell+{1\over 2}}\l({\pi s\over N}\r)
\end{equation}
for the Dirac fermion. We note again that for Majorana fermions, all the expressions for $K^f$ and $d_{\lambda\ell}$ would need to be divided by half.
\section{${1\over 4}$--BPS Black Holes in $\cn=4$ Supergravity}\label{sec5}
Having reviewed the essential ingredients for the computations that we will perform, we now turn to the main task at hand: computations of logarithmic corrections to exponentially suppressed saddle points of the quantum entropy function. We do this first for the case of ${1\over 4}$--BPS black holes in $\cn=4$ supergravity, for which we had studied the contribution of a vector multiplet in \cite{Gupta:2013sva} and found that the answer vanishes. As there is an entire class of $\cn=4$ string theories containing an arbitrary number of vector multiplets for which the corresponding microscopic answer vanishes, it was natural to conclude that the contribution of a single vector multiplet to the log term should vanish. We found that this was indeed the case in the macroscopic computation. In this section we will compute the contribution of the gravity multiplet to the log correction to black hole entropy. We find that this answer also vanishes once the zero modes of the graviton and gravitini are carefully accounted for. This result completes the matching of the macroscopic and microscopic results for $\cn=4$ black holes at the level of log terms.

Before describing the actual computation, we begin with an overview of the strategy and simplifications that will help us solve the problem. These remarks also carry over to the $\cn=8$ and $\cn=2$ results that we obtain in later sections. Firstly, the spectrum of quadratic fluctuations about the quarter--BPS black hole attractor geometry was completely analysed in \cite{Banerjee:2010qc,Banerjee:2011jp}. Imposing the $\zn$ orbifold \eqref{znorb} on the spectrum projects onto a subset of the modes. In particular, it does not change the eigenvalues of the kinetic operator, it only changes the degeneracy of the eigenvalue. Further, the overall strategy of \cite{Banerjee:2010qc,Banerjee:2011jp} was to study the heat kernel of fields minimally coupled to background gravity, include a coupling to the graviphoton flux and see how the system changes. It was found that while the eigenvalues shifted from those of fields minimally coupled to gravity, the degeneracy of the eigenvalues did not change. In particular, the degrees of freedom in quadratic fluctuations always organised themselves into scalars and Dirac fermions with shifted eigenvalues. In our computations we will therefore adopt the overall strategy of \cite{Banerjee:2010qc,Banerjee:2011jp}. In particular, we will take the spectrum obtained on the full attractor geometry and impose the $\zn$ orbifold by replacing the unquotiented degeneracies by the degeneracy \eqref{scalardeg} for the scalar and the degeneracy \eqref{fermiondeg} for the Dirac fermion. This leads to a further simplification in the problem. We had pointed out that the conical terms in the heat kernels \eqref{scalarads2s2} and \eqref{diracads2s2n} are finite in the $t\mapsto 0$ limit. Therefore, their contribution to the $t^0$ term in the heat kernel expansion is independent of the eigenvalues $E_{\lambda\ell}$ and remains the same even when the fields are coupled to the graviphoton flux. All the effects of the flux manifest themselves in the $t^0$ term contained in the unquotiented part of the heat kernels. This term has already been computed in \cite{Banerjee:2011jp}. To account for the fixed point contributions, we need to compute the finite $t^0$ terms, which are independent of the eigenvalues, and sum them up. We do this in the sections below.

Additionally, there is an additional discrete mode contribution from vectors, gravitini and the graviton, which will be computed explicitly as was done for the vector discrete modes in \cite{Gupta:2013sva}. The discrete modes also furnish zero modes to the kinetic operator, and these will also be separately accounted for. We finally find that the contribution to the log term from the gravity multiplet in $\cn=4$ supergravity also vanishes.
\subsection{Integer Spin Fields}
We begin with the contribution of the integer spin fields in the $\cn=4$ gravity multiplet to the one--loop determinant on $\adss\otimes\ss2$ with the graviphoton flux. The physical spectrum consists of one graviton, 6 gauge bosons $\ca^{(a)}$, $1\leq a\leq 6$, and two scalars which correspond to the axion-dilaton field. In addition, there are 12 scalar ghosts that arise during gauge-fixing the $U(1)$ gauge symmetries associated with $\ca^{(a)}$ and two vector ghosts which correspond to gauge fixing the group of linearised diffeomorphisms about the $\adss\otimes\ss2$ background. The vector modes also have a discrete series contribution which we shall compute.
We begin with the contribution of 4 vector fields $\ca^{(a)}$, $3\leq a\leq 6$ which do not couple to the background graviphoton flux. Their contribution on $\ads2s2n$ was evaluated in \cite{Gupta:2013sva}. It is given by
\begin{equation}
K^v=K^{(v_T,s)}+K^{(v_L,s)}+K^{(v_d,s)}+K^{(s,v_T)}+K^{(s,v_L)},
\end{equation}
where $T,L,d$ denote the transverse, longitudinal and discrete modes, and $K^{(v_T,s)}$ is the contribution of the transverse vector along the $\adss$ direction and the scalar along the $\ss2$ direction and so on. Now on the orbifolded space, we had evaluated $K^{(v_d,s)}$ in \cite{Gupta:2013sva} to be
\begin{equation}
K^{(v_d,s)}=-{1\over N}\sum_{s=0}^{N-1}\sum_{\ell=0}^\infty \chi_{\ell}\l({\pi s\over N}\r)e^{-{t\over a^2}\ell\l(\ell+1\r)}.
\end{equation}
Note that $\ell=0$ is a zero mode of the kinetic operator. We will account for the presence of these modes in our final results. We now turn to the contribution of the longitudinal and transverse modes of the 4 vector fields. Though the answer is already known from \cite{Gupta:2013sva} we will explicitly evaluate it again using the approach we have outlined above, i.e. by replacing the `unquotiented' degeneracy by the degeneracy on $\ads2s2n$. The contribution of the longitudinal and transverse modes of the 4 vector fields on $AdS_2\times S^2$ is given by \cite{Banerjee:2010qc,Banerjee:2011jp}\footnote{We have multiplied their results for the coincident heat kernel by the regularised volume of $\adss\otimes\ss2$, which is $-8\pi^2a^4$ to obtain these expressions.}
\begin{equation}
K^v_{U}=4\l[-\int_0^\infty d\lambda \lambda \tanh(\pi\lambda)\sum_{\ell=0}^\infty (2\ell+1)e^{-\frac{tE_{\lambda\ell}}{a^2}}(4-2\delta_{\ell,0})\r],
\end{equation}
where $E_{\lambda\ell}=\l(\lambda^2+{1\over 4}+\ell\l(\ell+1\r)\r)$. In what follows, we shall denote the heat kernel on the unquotiented space $\adss2\otimes\ss2$ by $K_U$, and the heat kernel on the $\zn$ orbifolded space $\ads2s2n$ as $K_N$. Thus we see that the entire heat kernel arranges itself into the scalar heat kernel where some $\ell=0$ modes are missing as they don't give rise to non-trivial gauge fields. The degeneracy of eigenvalues $E_{\lambda\ell}$ is given by
\begin{equation}
d_{\lambda\ell}=-\lambda\tanh\l(\pi\lambda\r)\l(2\ell+1\r).
\end{equation}
To obtain the expression for the heat kernel on the quotient space, we will replace this degeneracy with the new degeneracy \eqref{scalardeg}. We then obtain
\begin{equation}
\begin{split}
\tilde K^v_{N}(t)=\frac{1}{N}K^v_{U}(t)&+\frac{2}{N}\sum_{s=1}^{N-1}\sum_{\ell=0}^\infty \int_0^\infty d\lambda \chi_{\lambda}^b\l(\frac{\pi s}{N}\r)\chi_{\ell}\l(\frac{\pi s}{N}\r)e^{-\frac{E_{\lambda\ell}}{a^2}}(4-2\delta_{l,0})\\&-\frac{4}{N}\sum_{s=1}^{N-1}\sum_{\ell=0}^\infty\chi_{\ell}\l(\frac{\pi s}{N}\r)e^{-\frac{t\ell(\ell+1)}{a^2}}.
\end{split}
\end{equation}
As we are interested in the short-$t$ expansion of this heat kernel, in particular in the $\mathcal{O}\l(t^0\r)$ term, we will retain just the leading term in the short time expansion of the conical contributions. We therefore obtain
\begin{equation}
K^v_{N}=\frac{1}{N}K^v_{U}+\frac{2}{N}\sum_{s=1}^{N-1}\sum_{\ell=0}^\infty\int_0^\infty d\lambda \chi_{\lambda}^b\l({\pi s\over N}\r)\chi_{\ell}\l(\frac{\pi s}{N}\r)(4-2\delta_{l,0})-\frac{4}{N}\sum_{s=1}^{N-1}\sum_{\ell=0}^\infty\chi_{\ell}\l(\frac{\pi s}{N}\r).
\end{equation}
We can explicitly do the sums and integrals using the results listed in Appendix \ref{usefulsums} to obtain
\begin{equation}\label{kfreevec}
K^v_{N}={1\over N}K^v_{U} +2 {N^4-20N^2+19\over 45 N}+\mathcal{O}\l(t\r).
\end{equation}
Now due to the graviphoton flux, the graviton modes mix with those of the two gauge fields $\ca^{(1,2)}$ and the axion--dilaton scalars, but does not mix with any other fields \cite{Banerjee:2011jp}. We will now compute the heat kernel for the kinetic operator on this subset of fluctuations over $\ads2s2n$. Firstly, the entire kinetic operator for this system can be expressed in terms of modes of scalars on $\adss\otimes\ss2$ \cite{Banerjee:2011jp}. The heat kernel for this sub--system is given by a sum of two terms, the first of which corresponds to all scalar modes $\lambda$ along $\adss$ tensored with scalar modes on $\ss2$ with $\ell\geq 1$. The integrated heat kernel is given by equation 5.16 of \cite{Banerjee:2011jp}
\begin{equation}
\begin{split}
K^{g+2v}_{1,U}=&-e^{-{\bs\over 4}}\int_0^\infty d\lambda\,\lambda\tanh\l(\pi\lambda\r)e^{-\bs\lambda^2}\l[\sum_{\ell=1}^\infty\l(2\ell+1\r)e^{-2\bs\ell \l(\ell+1\r)}\l\lbrace 2 + 2e^{2\bs} +2e^{-2\bs}-2e^{2\bs}\delta_{\ell,1}\r.\r.\\&+ \l.\l.e^{i\bs\sqrt{2\lambda^2+2\ell\l(\ell+1\r)+{1\over 2}}} + e^{-i\bs\sqrt{2\lambda^2+2\ell\l(\ell+1\r)+{1\over 2}}} + e^{\bs\sqrt{2\lambda^2+2\ell\l(\ell+1\r)+{1\over 2}}} + e^{-\bs\sqrt{2\lambda^2+2\ell\l(\ell+1\r)+{1\over 2}}} \r.\r.\\ &+\l.\l.2e^{\bs\sqrt{2\ell\l(\ell+1\r)-2\lambda^2-{1\over 2}}}+ 2e^{-\bs\sqrt{2\ell\l(\ell+1\r)-2\lambda^2-{1\over 2}}}+\sum_{i=1}^6e^{-\bs f_{i}\l(\lambda,\ell\r)}\r\rbrace\r].
\end{split}
\end{equation}
Thus we see quite explicitly that the heat kernel is expressible in terms of scalar degrees of freedom. The eigenvalues are shifted from the $E_{\lambda\ell}=\l(\lambda^2+{1\over 4}+\ell\l(\ell+1\r)\r)$ in the absence of the graviphoton flux, but the degeneracy does not change from 
\begin{equation}
d_{\lambda\ell}=-\lambda\tanh\l(\pi\lambda\r)\l(2\ell+1\r).
\end{equation}
As argued previously, the effect of the orbifold is to replace the old degeneracy by \eqref{scalardeg}. We therefore obtain
\begin{equation}\label{kg2va}
\begin{split}
\tilde K^{g+2v}_{1,N}&=\frac{1}{N}K^{g+2v}_{1,U}+\frac{1}{2N}\sum_{s=1}^{N-1}\sum_{l=1}^\infty\int_0^\infty d\lambda\chi_{\lambda}^b\l(\frac{\pi s}{N}\r)\chi_l\l(\frac{\pi s}{N}\r)\l[20-2\delta_{l,1}\r]+\mathcal{O}\l(\bs\r)\\
%&=\frac{1}{N}K^{'m+2v}_{old}+\frac{1}{2N}\sum_{s=1}^{N-1}\frac{1}{2\sin^2\l(\frac{\pi s}{N}\r)}\l[20\sum_{l=1}^\infty \chi_l\l(\frac{\pi s}{N}\r)-2\chi_1\l(\frac{\pi s}{N}\r)\r]\\&=\frac{1}{N}K^{'m+2v}_{old}+\frac{1}{2N}\sum_{s=1}^{N-1}\frac{1}{2\sin^2\l(\frac{\pi s}{N}\r)}\l[20\sum_{l=0}^\infty \chi_l\l(\frac{\pi s}{N}\r)-20-2\chi_1\l(\frac{\pi s}{N}\r)\r]\\
&=\frac{1}{N}K^{'g+2v}_{U}+\frac{N^4+10N^2-11}{18N}-\frac{11(N^2-1)}{6N}+\frac{(N-1)(5-N)}{3N}+\mathcal{O}\l(\bs\r).
\end{split}
\end{equation}
Now we consider the $\ell=0$ modes of this subsystem. Their contribution to the heat kernel on the unquotiented space is given by equation 5.22 of \cite{Banerjee:2011jp}
\begin{equation}
K^{g+2v}_{2,U}=\int_0^\infty d\lambda \l(-\lambda\tanh\pi\lambda\r)e^{-\bs\l(\lambda^2+{1\over 4}\r)}\l[2+2e^{-2\bs}+e^{i \bs\sqrt{2\lambda^2+{1\over 2}}} + e^{-i \bs\sqrt{2\lambda^2+{1\over 2}}} +\sum_{i=1}^4 e^{-\bs g_i\l(\lambda\r)}\r].
\end{equation}
We recognise this as the heat kernel of 10 real scalars with eigenvalues shifted from the minimally coupled ones. We impose the $\zn$ orbifold by replacing the degeneracy of eigenvalues
\begin{equation}
d_{\lambda\ell}=-\lambda\tanh\pi\lambda \l(2\ell+1\r)
\end{equation}
for $\ell=0$ by the degeneracy \eqref{scalardeg} to obtain
\begin{equation}
K^{g+2v}_{2,U}={1\over N}K^{g+2v}_{1,U}+{10\over 2N}\sum_{s=1}^{N-1}\int_0^\infty d\lambda \chi_{\lambda}^b\l({\pi s\over N}\r)+\mathcal{O}\l(t\r),
\end{equation}
where again we exploited the fact that the fixed point contributions are finite in the $t\mapsto 0$ limit. This expression reduces to
\begin{equation}\label{kg2vb}
K^{g+2v}_{2,N}={1\over N}K^{g+2v}_{2,U}+5\l({N^2-1\over 6N}\r)+\mathcal{O}\l(t\r).
\end{equation}
Thus the complete contribution is obtained by adding \eqref{kg2va} and \eqref{kg2vb},
\begin{equation}\label{kg2v}
K^{g+2v}_{N}=\frac{1}{N}K^{m+2v}_{U}+\frac{N^4-14 N^2+36 N-23}{18 N}.
\end{equation}
Now we will consider the contribution coming from the mixing of the discrete mode of the vector field with the discrete modes of the graviton. We begin with a review of the modes themselves. Firstly, the two gauge fields $\ca^{(1)}$ and $\ca^{(2)}$ have discrete modes $v$ along the $\adss$ direction\footnote{The $m$s denote $\adss$ indices and the $\alpha$s denote $\ss2$ indices.} which furnish a basis\footnote{The basis is overcomplete as $v_m$ and $\epsilon_{mn}v^n$ are treated as independent vectors, this is compensated for by multiplying a factor of half to the heat kernel \cite{Banerjee:2011jp}.}
\begin{equation}\label{ads2s2discrete1}
\ca^{(1)}_m= E_1 v_m +\tilde{E}_1 \epsilon_{mn}v^n,\quad \ca^{(2)}_m= E_2 v_m +\tilde{E}_2 \epsilon_{mn}v^n,
\end{equation}
while the metric has the following discrete modes associated with it
\begin{equation}\label{ads2s2discrete2}
h_{m\alpha}={1\over\sqrt{\kappa_1}}\l(E_3\partial_\alpha v_m +\tilde{E}_3 \epsilon_{mn}\partial_\alpha v^n+ E_4\epsilon_{\alpha\beta}\partial^\alpha v_m +\tilde{E}_4 \epsilon_{\alpha\beta}\epsilon_{mn}\partial^\beta v^n\r),
\end{equation}
and 
\begin{equation}\label{ads2s2discrete3}
h_{mn}={a\over\sqrt{2}}\l(D_m\hat{\xi}_n+D_n\hat{\xi}_m-g_{mn}D^p\hat{\xi}_p\r),\quad \hat{\xi}_m=E_5 v_m +\tilde{E}_5 \epsilon_{mn}v^n.
\end{equation}
Also, there is an additional set of discrete modes given by
\begin{equation}\label{ads2s2discrete4}
h_{mn}=E_6 w_{mn},
\end{equation}
where $w_{mn}$ are the modes \eqref{metricdiscrete}. It was shown in \cite{Banerjee:2011jp} that the $E_5$ and $E_6$ modes don't mix with anything, while the rest of the modes mix with each other. Again, we first consider the modes with $\ell\geq 1$ along the $\ss2$ direction. We define $K^{discrete}_{1}$ to be the heat kernel of the $\ell\geq 1$ modes arising from the rediagonalisation of all the $E$ and $\tilde{E}$ modes except $E_6$. On the unquotiented space, it is given by (see equation (5.27) of \cite{Banerjee:2011jp})
\begin{equation}
\begin{split}
K^{discrete}_{1,U}=-\sum_{\ell=1}^\infty & \l(2\ell+1\r)e^{-\bs\ell\l(\ell+1\r)}\l\lbrace e^{-2\bs}+e^{-\bs\sqrt{2\ell\l(\ell+1\r)}}+e^{\bs\sqrt{2\ell\l(\ell+1\r)}}\r.\\&+\l.e^{i\bs\sqrt{2\ell\l(\ell+1\r)}}+e^{-i\bs\sqrt{2\ell\l(\ell+1\r)}}\r\rbrace.
\end{split}
\end{equation}
We recognize this as the heat kernel of 5 discrete vector modes with eigenvalues shifted from the Hodge Laplacian's $E_{\ell}={\ell\l(\ell+1\r)\over a^2}$ due to the mixing of the modes via the graviphoton flux. To compute the heat kernel of this subsystem on the $\zn$ orbifold we need to identify the degeneracy of the eigenvalue $E_{\ell}$ on $\ads2s2n$. This may be read off from the equation (5.29) of \cite{Gupta:2013sva}
\begin{equation}
K=-{1\over N}\sum_{s=0}^\infty\sum_{\ell=0}^\infty \chi_{\ell}\l({\pi s\over N}\r)e^{-{t\over a^2}\ell\l(\ell+1\r)},
\end{equation}
where we have included the contribution of the zero mode. This leads to the following identification for the degeneracy of an eigenvalue $E_\ell$ of a vector discrete mode on $\ads2s2n$.
\begin{equation}
d_{v,d}\l(\ell\r)=-{1\over N}\sum_{s=0}^{N-1}\chi_{\ell}\l({\pi s\over N}\r).
\end{equation}
Then the heat kernel of this subsystem on the $\zn$ orbifold is given by
\begin{equation}
K^{discrete}_{1,N}={1\over N}K^{discrete}_{1,U}-{5\over N}\sum_{s=1}^{N-1}\sum_{\ell=1}^\infty\chi_{\ell}\l({\pi s\over N}\r)+\mathcal{O}\l(t\r),
\end{equation}
which may be evaluated to obtain
\begin{equation}\label{kdiscreteboson1}
K^{discrete}_{1,N}={1\over N}K^{discrete}_{1,U}-5 {\l(N-1\r)\l(N-5\r)\over 6N}+\mathcal{O}\l(t\r).
\end{equation}
We now consider the contribution of the $\ell=0$ modes on the $\ss2$ to the discrete series heat kernel. The unorbifolded answer is \cite{Banerjee:2011jp}
\begin{equation}
K^{discrete}_{2,U}=-\l(2\cdot \ell+1\r)\l(2+e^{-2\bs}\r)\vert_{\ell=0}.
\end{equation}
On the orbifold space, the corresponding answer is given by
\begin{equation}
K^{discrete}_{2,N}={1\over N}K^{discrete}_{2,U}-{2\over N}\sum_{s=1}^{N-1} \chi_{0}\l({\pi s\over N}\r) +\mathcal{O}\l(t\r),
\end{equation}
which evaluates to
\begin{equation}\label{kdiscreteboson2}
K^{discrete}_{2,N}={1\over N}K^{discrete}_{2,U}-3\l({N-1\over N}\r)+\mathcal{O}\l(t\r).
\end{equation}
The final discrete mode contribution comes from the mode $E_6$ of the metric. The degeneracy of the eigenvalue $E_{\ell}$ labelled by the quantum number $\ell$ has been evaluated in \eqref{degE6l0} and \eqref{degE6l}. Using those expressions we obtain
\begin{equation}\label{kdiscreteboson3}
K^{discrete}_{3,N}={1\over N}K^{discrete}_{3,U} -\frac{(N-3) (N-1)}{2 N} +\mathcal{O}\l(t\r).
\end{equation}
The total contribution of discrete modes to the bosonic fields is given by the sum of \eqref{kdiscreteboson1}, \eqref{kdiscreteboson2} and \eqref{kdiscreteboson3}. We thus obtain
\begin{equation}\label{kdiscreteboson}
K^{discrete}_{N}={1\over N}K^{discrete}_{U}-4+{4\over 3N}\l(6N-N^2-2\r)+\mathcal{O}\l(t\r).
\end{equation}
We will now compute the ghost contribution to the determinants. The general form of the heat kernel is
\begin{equation}
K^{ghost}=-\sum_{\lambda,\ell}d_{\lambda\ell}e^{-\bs E_{\lambda\ell}},
\end{equation}
where the minus sign is on account of fermionic statistics of ghosts. We now evaluate these determinants. Firstly there are 12 scalar $U(1)$ ghosts each of which contributes the heat kernel of a free scalar \cite{Banerjee:2011jp}, with an overall minus sign. We therefore obtain
\begin{equation}\label{kghostb1}
K^{ghost}_{N}={1\over N}K^{ghost}_{U} - {3\over 2}{N^4+10N^2-11\over 45 N}+\mathcal{O}\l(t\r).
\end{equation}
Secondly, there is a contribution from vector ghosts arising from gauge fixing the diffeomorphism group. As they are vector modes, they can be classified into longitudinal and transverse modes, with an additional set of discrete modes. We first concentrate on the longitudinal and transverse modes. It was again shown explicitly in \cite{Banerjee:2011jp} that the heat kernel of these modes organised itself into the heat kernel over scalar degrees of freedom with shifted eigenvalues. As in \cite{Banerjee:2011jp}, we will compute the contribution of these modes in two sets. The first set are the modes with $\ell\geq 1$ along the $\ss2$ direction, and the second set are the modes with $\ell=0$. From equation (5.36) of \cite{Banerjee:2011jp}, we find that the degrees of freedom for the $\ell\geq 1$ modes correspond to 8 scalars with shifted eigenvalues, while for the $\ell=0$ modes they correspond to 4 scalars with shifted eigenvalues. This heat kernel may be evaluated exactly in the same manner as the computations carried out above. We finally find that the contribution of these modes to the ghost determinant is given by the heat kernel
\begin{equation}\label{kghostb2}
K^{ghost}_{2,N}={1\over N}K^{ghost}_{2,U}-\frac{N^4-5 N^2+4}{45 N}+\mathcal{O}\l(t\r).
\end{equation}
Finally, there is the contribution to the diffeomorphism ghosts from the discrete modes for vector fields on $\adss$ which tensor with eigenstates of the scalar Laplacian on $\ss2$. From equation (5.37) of \cite{Banerjee:2011jp}, we recognize this as the contribution of two discrete modes of vector fields on $\adss\otimes\ss2$ with shifted eigenvalues. On the quotient space this therefore becomes,
to order $t^0$,
\begin{equation}\label{kghostb3}
K^{ghost}_{3,N}={1\over N}K^{ghost}_{3,U}+{N^2-1\over 3N}+\mathcal{O}\l(t\r).
\end{equation}
The total ghost contribution is obtained by adding \eqref{kghostb1},\eqref{kghostb2} and \eqref{kghostb3} to obtain
\begin{equation}\label{kghostb}
K^{ghost}_{N}={1\over N}K^{ghost}_{U}-\frac{\left(N^2-1\right)^2}{18 N}+\mathcal{O}\l(t\r)
\end{equation}
The total bosonic contribution is given by the sum of the heat kernels of the 4 vector fields uncoupled to the graviphoton flux \eqref{kfreevec}, the metric coupled to the two vector fields $\ca^{(1,2)}$ and the axion--dilaton scalars \eqref{kg2v} and \eqref{kdiscreteboson}, and the total ghost contribution \eqref{kghostb}. We find, on adding these together, the total contribution of all the integer spin fields.
\begin{equation}\label{kb}
K^{B}_{N}={1\over N}K^{B}_{U}+2\l({N^4-65N^2+135N-71\over 45 N}\r)+\mathcal{O}\l(t\r).
\end{equation}
%%%%%%%%%%%%%%%%%%%%%%%%%%%%%%%%%%%%%%%%%%%%%%%%
\subsection{Half--Integer Spin Fields}
We now turn to studying the contribution to the heat kernel on the $\zn$ orbifold \eqref{znorb} of the black hole attractor geometry from fields of half--integer spin. The physical fields are four gravitino fields and four dilatino fields. In addition, there are fermionic ghosts as well. These arise from gauge--fixing supersymmetry. Just as the determinants over the integer--spin fields arranged themselves into determinants over scalar modes, it may be shown that determinants over half--integer spins arrange themselves into determinants over Dirac fermions \cite{Banerjee:2011jp}. We now compute the contributions of all such fields to the one-loop determinant. We first concentrate on the contribution of the $(\lambda,\ell)$ modes, where $\ell\geq 1$, of the spin-${3\over 2}$ and spin-${1\over 2}$ fields which mix with each other due to the graviphoton flux. While the full mixing matrix is quite complicated, we don't need it explicitly here. We will again use the fact that the effect of the graviphoton flux is to shift the eigenvalues, but the degeneracies don't change. It was explicitly shown in \cite{Banerjee:2011jp} that the degrees of freedom in this subset of fluctuations are equal to those of 10 Dirac fermions on $\adss\otimes\ss2$. Then, by the arguments presented above, the heat kernel over the $\l(\lambda,\ell\geq 1\r)$ modes of the fermions on $\ads2s2n$ is given by
\begin{equation}
K^{f}_{1,N}={1\over N}K^f_{1,U}-10\cdot {2\over N}\sum_{s=1}^{N-1}\sum_{\ell=1}^\infty\int_0^\infty d\lambda\,\chi_{\,\lambda,\ell+{1\over 2}}^{f}\l({\pi s\over N}\r)+\mathcal{O}\l(t\r).
\end{equation}
Note that we are now considering the fermion heat kernel with an overall minus sign. This is on account of the Grassmann nature of fermions. This expression evaluates to
\begin{equation}\label{kf1}
K^{f}_{1,N}={1\over N}K^f_{1,U}-{N^4-65N^2+180N-116\over 9N}\mathcal{O}\l(t\r).
\end{equation}
Similarly, from  equation 6.16 of \cite{Banerjee:2011jp}, we see that the $\ell=0$ modes along the $\ss2$ have the degrees of freedom of 8 Dirac fermions. We therefore have
\begin{equation}
K^{f}_{2,N}={1\over N}K^f_{2,U}-8\cdot {2\over N}\sum_{s=1}^{N-1}\int_0^\infty d\lambda\,\chi_{\,\lambda,{1\over 2}}^{f}\l({\pi s\over N}\r)+\mathcal{O}\l(t\r).
\end{equation}
This can be evaluated to obtain
\begin{equation}\label{kf2}
K^{f}_{2,N}={1\over N}K^{f}_{2,U}-16\l({\l(N-1\r)\l(N-2\r)\over 3N}\r)+\mathcal{O}\l(t\r).
\end{equation}
In addition to the modes of the spin--$3\over 2$ field $\Psi_{\m}$ labelled by the quantum numbers $\l(\lambda,\ell\r)$ there is an additional discrete series where the vector leg of $\Psi_\m$ points along the $\adss$ direction. This corresponds to choosing $\lambda=i$ in the $\Psi_\m$ modes. These fields also contribute to the one-loop determinant and we evaluate the contribution of these modes now. From Equation 6.22 of \cite{Banerjee:2011jp} we see that the degrees of freedom contributing to the determinant are those of two such fields. Then the heat kernel over the discrete modes can be evaluated using the degeneracies computed in Appendix \ref{discretesec}. We finally obtain
\begin{equation}\label{kfdisc}
K^{f,discrete}_{N}={1\over N}K^{f,discrete}_{U}+4\l({N^2-3N+2\over 3N}\r)+\mathcal{O}\l(t\r).
\end{equation}
We now are left with the contribution of the ghost determinant. From equation (6.23) of \cite{Banerjee:2011jp} we see that the degrees of freedom are those of 6 Dirac fermions. 
\begin{equation}\label{kfghost}
K^{ghost}_{N}={1\over N}K^{ghost}_{U}+{N^4-5N^2+4\over 15 N}+\mathcal{O}\l(t\r).
\end{equation}
We note that this expression has been evaluated with an overall minus with respect to the other fermion heat kernels, this is because fermionic ghosts are Grassmann even. The total contribution of fermions to the heat kernel is then given by adding \eqref{kf1}, \eqref{kf2}, \eqref{kfdisc} and \eqref{kfghost}. We thus obtain
\begin{equation}\label{kf}
K^{F}_{N}={1\over N}K^{F}_{U}-2\l({N^4-65N^2+180N-116\over 45N}\r)+\mathcal{O}\l(t^0\r).
\end{equation}
\subsection{The Logarithmic Correction}
The total heat kernel on the $\zn$ orbifold background is the sum of \eqref{kb} and \eqref{kf}. It is given by
\begin{equation}
K^{total}_{N}={1\over N} K^{total}_{U}-2+{2\over N}+\mathcal{O}\l(t\r).
\end{equation}
Both the heat kernels $K^{total}_{new}$ and $K^{total}_{old}$ have been evaluated over the full set of modes of the theory, including non-zero modes. We now focus on the $t^0$ term in the heat kernel expansion, as this is what contributes to the log term. On the unquotiented space the bosonic contribution to $K^{log}$ was evaluated in Equation 5.43 of \cite{Banerjee:2011jp}. It was found that the contribution to the coincident heat kernel, where the zero modes were not subtracted out, was given by\footnote{The expression in \cite{Banerjee:2011jp} is for the coincident heat kernel on $\adss\otimes\ss2$. To obtain the corresponding contribution to the integrated heat kernel, we have multiplied it by $-8\pi^2a^4$, the regularized volume of $\adss\otimes\ss2$.}
\begin{equation}
K^B_{U}\l(0;t\r)=-{338\over 45}.
\end{equation}
Additionally, the fermion contribution to the integrated heat kernel, without subtracting out zero modes, was evaluated in Equation 6.33 of \cite{Banerjee:2011jp} to be
\begin{equation}
K^F_{U}\l(0;t\r)={248\over 45}.
\end{equation}
Then the total contribution to the $t^0$ term of the heat kernel on the unquotiented space is given by the sum of the two terms above. We therefore obtain
\begin{equation}
K_{N}\l(0;t\r)=-{2\over N}-2+{2\over N}=-2.
\end{equation}
From this we can extract the term proportional to $\ln\l(a\r)$ in the partition function using \eqref{logcontribution}. We need the coefficients $\beta$ for the vector, spin--${3\over 2}$, and graviton fields respectivelt \cite{Banerjee:2010qc,Banerjee:2011jp}
\begin{equation}
\beta_v=1,\quad\beta_{3\over 2}=3,\quad \beta_{g}= 2.
\end{equation}
Further, the number of zero modes of the graviton is 
\begin{equation}
n^0_g=-2,
\end{equation}
as there are two zero modes of the metric, one from the $E_6$ modes and the other from the mixing of the vector and graviton discrete modes. Each of these contributes -1 to the number of zero modes. Further, the number of fermion zero modes is 
\begin{equation}
n^0_{3\over 2}=2,
\end{equation}
where the plus sign is due to our sign conventions for the fermion heat kernel. We then obtain the total contribution to the log of the quantum entropy function proportional to $\ln\l(a\r)$ when expanded about the saddle point \eqref{attractor}, \eqref{orbact} to be
\begin{equation}
\ln\l(d_{hor,N}\r)_{\log}=K_{N}\l(0;t\r)+n_{3/2}\l(\beta_{3\over 2}-1\r)+n_g\l(\beta_g-1\r)=0.
\end{equation}
We thus see that the overall log contribution of the gravity multiplet in $\cn=4$ theories also vanishes, completing the proof of vanishing log terms in $\cn=4$ string theory.
%%%%%%%%%%%%%%%%%%%%%%%%%%%%%%%%%%%%%%%%%%%%%%%%%%%%%
\section{${1\over 8}$--BPS black holes in $\cn=8$ Supergravity}\label{sec6}
We now extend the computation of section \ref{sec4} for the quarter--BPS black hole in $\cn=4$ supergravity to the case of the $1\over 8$--BPS black holes in $\cn=8$ supergravity. Following \cite{Banerjee:2011jp} we will use the fact that there is a consistent truncation of $\cn=8$ supergravity to $\cn=4$ supergravity. Then the quarter BPS black hole of $\cn=4$ supergravity can be regarded as the ${1\over 8}$ BPS black hole in $\cn=8$ supergravity. Then, the one-loop determinant in $\cn=8$ supergravity recieves contributions from the $\cn=4$ fields and also from the extra $\cn=8$ fields. The contribution from the $\cn=4$ fields has already been shown to vanish. We will concentrate on the extra $\cn=8$ fields.
\subsection{Bosonic Contribution}
We now compute the contribution of the extra bosonic fields. These are 16 gauge fields and 32 scalars, which we denote by $\ca_\m^{(r)}$, $\phi_{1(r)}$ and $\phi_{2(r)}$ respectively, where $r\in\l[1,16\r]$. For $r\in\l[1,8\r]$ the gauge field along $\ss2$ couples to the $\phi_{2(r)}$ via the background flux, and for $r\in\l[9,16\r]$ the gauge field along $\adss$ couples to the $\phi_{1(r)}$ via the background flux. Therefore, firstly we have 16 free scalars $\phi_{1(r)}$, $r\in\l[1,8\r]$ and $\phi_{2(r)}$, $r\in\l[9,16\r]$. These contribute
\begin{equation}\label{kb1}
K^b_{1,N}={1\over N}K^b_{1,U}+2\l({N^4+10N^2-11\over 45N}\r)+\mathcal{O}\l(t\r).
\end{equation}
We now consider how the gauge field mixes with the scalar along the $\ss2$ direction. The heat kernel for this system is given by
\begin{equation}
K^b_{2}=K^{(v,s)}+K^{(s,s+v)}-2K^{(s,s)},
\end{equation}
where $K^{(v,s)}$ is the heat kernel for the vector along $\adss$ and the scalar along $\ss2$, $K^{(s,s+v)}$ is the heat kernel for the scalar along $\adss$ and the vector-scalar mixed system along $\ss2$. The third term is the ghost determinant. This further decomposes into
\begin{equation}
K^b_2=K^{(v_T,s)}+K^{(v_L,s)}+K^{(v_d,s)}+K^{(s,s+v_T)}+K^{(s,v_L)}-2K^{(s,s)},
\end{equation}
where we have further decomposed the gauge field into longitudinal, transverse and discrete modes. There is no mixing with the scalar for these modes. The contribution of the discrete modes has already been computed in \cite{Gupta:2013sva}
\begin{equation}\label{kvds}
K^{(v_d,s)}_{\ads2s2n}=-K^s_{\sn}={1\over N}K^{(v_d,s)}_{old}-{N^2-1\over 6N}+\mathcal{O}\l(t\r).
\end{equation}
Also the heat kernel for the transverse or longitudinal vector along $\adss$ is just the heat kernel for the scalar \cite{Banerjee:2010qc,Banerjee:2011jp}, i.e. $K^{(v_T,s)}=K^{(v_L,s)}=K^{(s,s)}$. We therefore have 
\begin{equation}
K^{(s,v_L)}_{N}={1\over N}K^{(s,v_L)}_{U}+{1\over 2N}\sum_{s=1}^{N-1}\sum_{\ell=1}^\infty\int_{0}^\infty d\lambda\, \chi_{\lambda,\ell}^b\l({\pi s\over N}\r)+\mathcal{O}\l(t\r).
\end{equation}
This may be evaluated to obtain
\begin{equation}\label{ksvl}
K^{(s,v_L)}_{N}={1\over N}K^{(s,v_L)}_{U}+{N^4+10N^2-11\over 360N}-{N^2-1\over 12N}+\mathcal{O}\l(t\r). 
\end{equation}
Now eigenvalues shift in the heat kernel $K^{(s,s+v_T)}$, but as we have previously argued, the shift is irrelevant for computing the $t^0$ contribution from the conical terms.
\begin{equation}\label{kssvt}
K^{(s,s+v_T)}_{N}={1\over N}K^{(s,s+v_T)}_{U}+{1\over N}\sum_{s=1}^{N-1}\sum_{\ell=1}^\infty\int_{0}^\infty d\lambda\, \chi_{\lambda,\ell}^b\l({\pi s\over N}\r)+\mathcal{O}\l(t\r).
\end{equation}
Now we add up the equations \eqref{kvds},\eqref{ksvl},\eqref{kssvt} with the heat kernel for free scalars on $\ads2s2n$, and multiply by 8 because there are 8 such vector-scalar mixed systems, and obtain 
\begin{equation}\label{kb2}
K^b_{2,new}={1\over N}K^b_{2,old}+\frac{N^4-30 N^2+29}{15 N}+\mathcal{O}\l(t\r).
\end{equation}
Similarly, we can compute the heat kernel for the 8 vector--scalar system where the vector mixes with the scalar along the $\adss$ direction. The final answer is given by
\begin{equation}\label{kb3}
K^b_{3,N}={1\over N}K^b_{3,U}+\frac{N^4-30 N^2+29}{15 N}+\mathcal{O}\l(t\r).
\end{equation}
The full bosonic contribution is the sum of \eqref{kb1}, \eqref{kb2} and \eqref{kb3} and is given by
\begin{equation}\label{kbn8}
K^{b}_{N}={1\over N}K^{b}_{U}+8\l({N^4-20N^2+19\over 45N}\r)+\mathcal{O}\l(t\r).
\end{equation}
\subsection{Fermionic Contribution}
First we have 4 free fermion fields, each of which is equivalent to 4 Majorana fermions each. Therefore, this subset has the degrees of freedom of 8 Dirac fermions in total \cite{Banerjee:2011jp}. Their contribution is
\begin{equation}
K^f_{1,N}={1\over N}K^f_{1,U}-8\cdot{2\over N}\sum_{s=1}^{N-1}\sum_{\ell=0}^\infty\int_0^\infty d\lambda\,\chi^f_{\,\lambda,\ell+{1\over 2}}\l({\pi s\over N}\r)e^{-{t\over a^2}E_{\lambda\ell}}.
\end{equation}
This evaluates to
\begin{equation}\label{kf1n8}
K^f_{1,N}={1\over N}K^f_{1,U}-4\l({N^4-5N^2+4\over 45 N}\r)+\mathcal{O}\l(t\r).
\end{equation}
Now we turn to the rediagonalised fermions in the principal series. From equation (8.17) of \cite{Banerjee:2011jp} we see that the $\ell\geq 1$ modes have degrees of freedom corresponding to 14 Dirac fermions. Hence they contribute
\begin{equation}
K^f_{2,N}={1\over N}K^f_{2,U}-14\cdot{2\over N}\sum_{s=1}^{N-1}\sum_{\ell=1}^\infty\int_0^\infty d\lambda\,\chi^f_{\,\lambda,\ell+{1\over 2}}\l({\pi s\over N}\r)e^{-{t\over a^2}E_{\lambda\ell}},
\end{equation}
which may be evaluated to obtain
\begin{equation}\label{kf2n8}
K^f_{2,N}={1\over N}K^f_{2,U}-7\l({N^4-65N^2+180N-116\over 45 N}\r)+\mathcal{O}\l(t\r).
\end{equation}
Next, we compute the contribution of the $\ell=0$ modes. Equation (8.19) of \cite{Banerjee:2011jp} shows that the degrees of freedom correspond to the $\ell=0$ modes of 12 Dirac fermions. We therefore obtain
\begin{equation}
K^f_{3,N}={1\over N}K^f_{3,U}-12\cdot{2\over N}\sum_{s=1}^{N-1}\sum_{\ell=0}^\infty\int_0^\infty d\lambda\,\chi^f_{\,\lambda,\ell+{1\over 2}}\l({\pi s\over N}\r)+\mathcal{O}\l(t\r),
\end{equation}
which can be evaluated to obtain
\begin{equation}\label{kf3n8}
K^f_{3,N}={1\over N}K^f_{3,U}-\frac{8}{N}(N-1)(N-2)+\mathcal{O}\l(t\r).
\end{equation}
Next we will calculate the contribution of the discrete modes. Comparing with equation (8.24) of \cite{Banerjee:2011jp}, we see that this corresponds to the degrees of freedom contained in two discrete mode fields built out of Dirac fermions. Their contribution is therefore given by
\begin{equation}\label{kf4n8}
K^{f}_{4,N}=\frac{1}{N}K^{f}_{4,U}+\frac{4(N^2-3N+2)}{3N}+\mathcal{O}\l(t\r).
\end{equation}
Finally  we compute the ghost contribution. Equation (8.26) of \cite{Banerjee:2011jp} indicates that the degrees of freedom are those of 6 Dirac fermions. The ghost contribution is therefore given by
\begin{equation}\label{kf5n8}
K_{new}^{ghost}=\frac{1}{N}K_{old}^{ghost}+\frac{(N^4-5N^2+4)}{15N}+\mathcal{O}\l(t\r).
\end{equation}
Thus the complete fermonic contribution is given by the sum of \eqref{kf1n8}, \eqref{kf2n8}, \eqref{kf3n8}, \eqref{kf4n8}, \eqref{kf5n8}, and reads as
\begin{equation}\label{kfn8}
K^{F}_{N}=\frac{1}{N}K^{F}_{U}-\frac{8 (-26+45N-20N^2 + N^4)}{45 N}+\mathcal{O}\l(t\r).
\end{equation}
\subsection{The Logarithmic Correction}
We now use these results to compute the logarithmic correction to the quantum entropy function in $\cn=8$ supergravity. As the extra fields that we have considered in this section have no zero modes, the logarithmic correction only receives contributions from the $t^0$ term in the heat kernel expansion. Adding the bosonic and fermonic contributions \eqref{kbn8} and \eqref{kfn8}, we get for the order $t^0$ term
\begin{equation}
K_{N}\l(0;t\r)=\frac{1}{N}K_{U}\l(0;t\r)-8+\frac{8}{N}.
\end{equation}
Also, from \cite{Banerjee:2011jp}, $K_{old}\l(0;t\r)=-8$. Thus the logarithmic term on the orbifold space given by
\begin{equation}
\ln\l(d_{hor,N}\r)_{\log}=\l(-\frac{8}{N}-8+\frac{8}{N}\r)\ln a=-8\ln a =-4\ln\l({A_H\over G_N}\r),
\end{equation}
which is the same as the log correction in the unorbifolded case. This matches perfectly which the prediction from the microstate counting \cite{Sen:2009gy}.
\section{$1\over 2$--BPS Black Holes in $\cn=2$ Supergravity}\label{sec7}
We now turn to the corresponding computation of the log term about the $\zn$ orbifold of the attractor geometry in case of half--BPS black holes in $\cn=2$ supergravity. In case of the STU model, the corresponding computation was done about the leading saddle point in \cite{Banerjee:2011jp}, and subsequently the computation was carried out for a generic $\cn=2$ theory in \cite{Sen:2011ba}. This is the computation that we shall also follow. Again, our strategy will be to enumerate the degrees of freedom in the heat kernel and then impose the $\zn$ orbifold on them by changing the degeneracies to those computed on the orbifold. Details of the eigenvalues of the kinetic operator will only affect the contribution from the unquotiented answer, which has already been computed in \cite{Sen:2011ba}. As the computations are identical to those carried out in Sections \ref{sec5} and \ref{sec6}, we will mostly state the final results.

Following \cite{Sen:2011ba} we consider quadratic fluctuations in the near horizon geometry of a half--BPS black hole. The one-loop determinants in $\cn=2$ supergravity receive contributions from vector, hyper and gravity multiplets, which we now evaluate. Firstly, consider the contribution from a vector multiplet, which contains a vector field and two real scalars. It can be shown that the bosons couple to the background fields in the same way as the bosons in the $\cn=4$ vector multiplet, excepting the contribution of the four extra free scalars in the $\cn=4$ vector multiplet. The contribution of these fields was analysed in \cite{Gupta:2013sva} on the orbifold geometry and we quote the final result. The bosonic heat kernel is given by
\begin{equation}\label{kbv}
K^b_{N} = {1\over N} K^b_{U} + \frac{N^4-20 N^2+19}{90 N} +\mathcal{O}\l(t\r).
\end{equation}
The contribution of the ghost determinant has been included in this. The fermionic fields are two Majorana fermions, or equivalently, one Dirac fermion. These contribute
\begin{equation}\label{kfv}
K^f_{N} = {1\over N} K^f_{U} -{N^4-5N^2+4\over 90 N} +\mathcal{O}\l(t\r).
\end{equation}
Then we can add \eqref{kbv} and \eqref{kfv} to obtain the heat kernel of a single vector multiplet. Focusing on the $t^0$ term, we find 
\begin{equation}
K^{v.m.}_{N}\l(0;t\r)={1\over N} K^{v.m.}_{U}\l(0;t\r)-\frac{N^2-1}{6 N}.
\end{equation}
On using equation (6.7) of \cite{Sen:2011ba}, we finally obtain
\begin{equation}\label{kvm}
K^{v.m.}_{N}\l(0;t\r)= -{N\over 6}.
\end{equation}
Similarly, for a hypermultiplet, the bosonic degrees of freedom are of 4 real scalars minimally coupled to background gravity, while the fermionic degrees of freedom are those of one Dirac fermion. We therefore have
\begin{equation}\label{kbm}
K^{b}_{N}={1\over N} K^{b}_{U} + {N^4+10N^2-11\over 90 N}+ \mathcal{O}\l(t\r),
\end{equation}
and 
\begin{equation}
K^f_{N}={1\over N}K^f_{U}- {N^4-5N^2+4\over 90N} +\mathcal{O}\l(t\r).
\end{equation}
Therefore the $t$ independent term in the heat kernel of one hypermultiplet in the $\zn$ orbifold of the near--horizon geometry of the half--BPS black hole is given by
\begin{equation}\label{khm}
K^{h.m.}_{N}\l(0;t\r)=+{N\over 6},
\end{equation}
where we have also used equation (6.16) of \cite{Sen:2011ba} for the unquotiented heat kernel. We are finally left with computing the contribution of the gravity multiplet to the one-loop determinant. We will first compute the contribution from fields of integer spin. As always, we compute the heat kernel over all possible modes in the theory and take into account the presence of zero modes right at the end. The physical integer--spin fields in an $\cn=2$ vector multiplet are one graviton and one graviphoton. In addition, there are ghost contributions. We begin with computing the determinant over the graviton and graviphoton system. The degrees of freedom in the principal series arrange themselves into 14 scalar fields on $\adss\otimes\ss2$, out of which two scalar fields contribute when $\ell\geq 2$ and 6 contribute when $\ell\geq 1$. The heat kernel evaluated over these fields is given by
\begin{equation}\label{kgmb}
K^{b}_{N}={1\over N}K^{b}_{U} + \frac{7 N^4-140 N^2+360 N-227}{180 N}.
\end{equation}
We now turn to the contribution of the discrete modes. Again we only need to count degrees of freedom. These take the form of one $E_6$ mode of the graviton, discussed in Section \ref{sec5}, which does not mix with any other field and the rest of the discrete modes organise themselves into four vector discrete modes where $\ell\geq 1$ and two vector discrete modes with $\ell=0$. The eigenvalues are of course shifted due to the coupling to the background graviphoton flux, but this is not relevant to computing the order $t^0$ contribution from the fixed points. In that case the total discrete mode contribution evaluates to
\begin{equation}\label{kgmdb}
K^{d}_{N}= {1\over N}K^{d}_{U}-{7N^2-24N+17\over 6N}+\mathcal{O}\l(t\r).
\end{equation}
We now turn to the ghost contribution. Firstly, the ghosts arising during the gauge fixing of the $U(1)$ gauge invariances give a contribution of two scalars to the ghost determinant. Secondly, the vector ghosts responsible for gauge fixing diffeomorphisms give a principal series contribution of 8 scalar fields, out of which only four contribute with $\ell=0$. Finally, the vector ghosts also contribute with two discrete modes. The eigenvalues associated to these modes are again not relevant to computing the $t^0$ contribution from the fixed points. The ghost contribution to the heat kernel is then given by
\begin{equation}\label{kgmghb}
K^{ghost}_{N}={1\over N}K^{ghost}_{U} - {N^4-14N^2+13\over 36 N}+\mathcal{O}\l(t\r).
\end{equation}
The total contribution of the integer--spin fields is given by the sum of \eqref{kgmb}, \eqref{kgmdb} and \eqref{kgmghb} and takes the form
\begin{equation}\label{kgmbtot}
K^{B}_{N}={1\over N}K^{B}_{U} + \frac{N^4-140 N^2+540 N-401}{90 N}+\mathcal{O}\l(t\r).
\end{equation}
The second contribution to the heat kernel over the gravity multiplet is given by the fields of half--integer spin, also analysed in \cite{Sen:2011ba} for the unorbifolded case. Firstly, we will compute the contribution of the physical fields, the gravitini, which are two Majorana spinors $\phi_\m$ and $c_\m$. The principal series contribution to the heat kernel contains degrees of freedom corresponding to four Dirac fermions in $\adss\otimes\ss2$ contributing with $\ell\geq 1$ modes along the $\ss2$ direction and 3 Dirac fermions contributing $\ell=0$ modes, while the discrete series contribution corresponds to one discrete mode. Finally, the ghost fields contribute with 3 Dirac fermions. Then the total determinant computed over the half-integer spin fields is given by
\begin{equation}\label{kgmftot}
K^{F}_{N}={1\over N}K^{F}_{U} - {N^4-125N^2+360N-236\over 90 N} +\mathcal{O}\l(t\r).
\end{equation}
The total heat kernel over all modes in the gravity multiplet is then given by
\begin{equation}
K^{g.m.}_{N}={1\over N}K^{g.m.}_{U} -{N\over 6}-{11\over 6N}+2.
\end{equation}
Now adding equations (4.36) and (5.29) of \cite{Sen:2011ba} to obtain the total heat kernel in the unquotiented background\footnote{We multiply the resultant expression by $-8\pi a^2$, the regularised volume of $\adss\otimes\ss2$.} we find that
\begin{equation}
K^{g.m.}_{U}={11\over 6}.
\end{equation}
Hence, the heat kernel in the $\zn$ orbifolded background is given by
\begin{equation}
K^{g.m.}_{N}=2-{N\over 6}.
\end{equation}
Now we need to account for zero modes. By an analysis similar to that carried out in Section \ref{sec5}, one can show that the number of graviton zero modes is -2, while the number of gravitino zero modes is +2. Again, the sign of the gravitino zero mode is chosen keeping in mind that the field is Grassmann odd. Then, using \eqref{logcontribution}, the contribution to the log term is given by
\begin{equation}
\ln\l(d_{hor,N}\r)_{\log}^{g.m.}= 4-{N\over 6}.
\end{equation}
Further, using \eqref{kvm} and \eqref{khm}, we can write down the contribution to the log term in a theory with $n_H$ hypermultiplets and $n_V$ vector multiplets. It is given by
\begin{equation}
\ln\l(d_{hor,N}\r)_{\log}=\l(4+{N\over 6}\l(n_H-n_V-1\r)\r)\ln\l(a\r).
\end{equation}
Identifying $\chi=2\l(n_V-n_H+1\r)$ as the Euler characterestic of the Calabi-Yau that the II(A) string theory is compactified on to obtain the $\cn=2$ string theory, and trading in the $\adss$ radius for the area of the horizon, we obtain
\begin{equation}
\ln\l(d_{hor,N}\r)_{\log}=\l(2-{N\chi\over 24}\r)\ln\l({A_H\over G_N}\r).
\end{equation}
\section{Conclusions and Outlook}
In this paper we exploited the heat kernel techniques developed in \cite{Gupta:2013sva} to compute the next to leading term in the quantum entropy function, when expanded about an exponentially suppressed saddle point. These saddle points are expected to correspond to exponentially suppressed saddle points in the asymptotic formulae for the microscopic degeneracy of the corresponding extremal black hole in $\cn=4$ and $\cn=8$ supergravity. We found that our macroscopic results matched perfectly with the microscopic results obtained within string theory. This provides non-trivial evidence that these macroscopic and microscopic saddle point should indeed be identified to each other. We also extended our computation to the case of the half--BPS black hole in $\cn=2$ supergravity. Though here the corresponding microscopic result is currently not available we found that the answer organises itself naturally in terms of the Euler character of the Calabi Yau that the theory is compactified on. Additionally, the answer has a curious feature that the log term grows with $N$. This indicates that if $N$ is sufficiently large, i.e. of the order of $\sqrt{A_H\over G_N}$, the log correction can overwhelm the leading classical answer. It would be interesting to pursue this course further and explicate the physics of these saddle points, assuming they exist for such large values of $N$ in the quantum theory. Perhaps the techniques of localisation \cite{Dabholkar:2010uh,Dabholkar:2011ec,Gupta:2012cy} applied to these saddle points can help us answer this question. Finally, we note that an intriguing match was found between the log term computed about the leading saddle point in $\cn=2$ supergravity and that obtained from an asymptotic expansion of a microscopic formula proposed by \cite{Denef:2007vg} as a refinement of the original OSV conjecture \cite{Ooguri:2004zv}. It would be very interesting to explore the matching in context of these saddle points as well. This is work in progress and we hope to report on it soon.
\acknowledgments
We would like to thank Justin David, Rajesh Gopakumar and Ashoke Sen for several very helpful discussions and correspondence. SL's work is supported by National Research Foundation of Korea grants 2005-0093843, 2010-220-C00003 and 2012K2A1A9055280.
\appendix
\section*{Appendix}
\section{The Discrete Modes for the Graviton}\label{spin2zerosec}
In this section we will discuss the contribution from the discrete modes of the graviton to the heat kernel of the spin-2 Laplacian on $\ads2s2n$. Before doing this, we would like to remind the reader of the origin of these modes \cite{Camporesi:1994ga}. In order to do so, the discussion in \cite{Banerjee:2011jp} will be helpful. 

Our strategy will be to express vector and spin-2 modes on $\adss$ in terms of derivative operators acting on scalar fields. Given normalisable eigenfunctions of the scalar Laplacian on $\adss$, denoted by $W_\lambda$ respectively, we can construct a basis\footnote{We omit normalisation constants which may be found in \cite{Banerjee:2011jp}.} of vector fields on $\adss$ as \cite{Banerjee:2010qc,Banerjee:2011jp}
\begin{equation}
\partial_m W_\lambda, \quad \epsilon_{mn}\partial^n W_\lambda.
\end{equation}
These eigenfunctions belong to the eigenvalue $E_\lambda={1\over a^2}\l(\lambda^2+{1\over 4}\r)$. See \cite{Camporesi:1994ga,Banerjee:2010qc,Banerjee:2011jp} for details and explicit expressions. This does not exhaust all possible normalisable vector modes on $\adss$. In particular, there is an additional series of modes given by \cite{Camporesi:1994ga}
\begin{equation}\label{vectordiscrete}
A=d\Phi,\quad \Phi = {1\over\sqrt{2\pi\vert\ell\vert}}\l[{\sinh\eta\over 1+ \cosh\eta}\r]^{\vert\ell\vert}e^{i\ell\theta},\quad\ell=\pm 1,\pm 2,\pm 3,\cdots.
\end{equation}
Though $\Phi$ itself is not normalisable, the vector field $A$ is normalisable over $\adss$. Formally, these modes can be arrived at by setting $\lambda={i\over 2}$ in $W_\lambda$ above. We will refer to $W_{\lambda}$ for $\lambda\in\mathbb{R}_+$ and the gauge fields obtained from its derivatives as the `principal series', while $A$ is referred to as the `discrete series'. 

In a similar manner, a basis for a symmetric rank--two tensor is given by\cite{Banerjee:2011jp}
\begin{equation}
g_{mn}W_\lambda, \quad \l(D_m v_n +D_n v_m -g_{mn}D^pv_p\r),
\end{equation}
where $D_m$ is the $\adss$ covariant derivative, and $v_m$ is a vector field on $\adss$ which may be either in the principal or the discrete series. In addition there is a further set of discrete modes of the spin--2 field, given by \cite{Camporesi:1994ga}
\begin{equation}\label{metricdiscrete}
\begin{split}
h_{\ell,mn}dx^mdx^n ={a\over\sqrt{\pi}}&\l[{\vert\ell\vert\l(\ell^2-1\r)\over 2}\r]^{1\over 2} {\l(\sinh\eta\r)^{\vert\ell\vert-2}\over\l(1+\cosh\eta\r)^{\vert\ell\vert}}e^{i\ell\theta} \l(d\eta^2+2i\sinh\eta d\eta d\theta -\sinh^2\eta d\theta^2\r),\\& \ell\in\mathbb{Z},\,\vert\ell\vert\geq 2.
\end{split}
\end{equation}
The modes \eqref{vectordiscrete} and \eqref{metricdiscrete} tensor with modes on the $\ss2$ to give a discrete series of modes on $\adss2\otimes\ss2$, enumerated in equations \eqref{ads2s2discrete1}, \eqref{ads2s2discrete2}, \eqref{ads2s2discrete3}, and \eqref{ads2s2discrete4}. We already computed the heat kernel over the vector discrete modes in \cite{Gupta:2013sva}. We will do the corresponding computation for the $h_{\ell,mn}$ modes of equation \eqref{metricdiscrete}. This is nothing but the $E_6$ mode of \eqref{ads2s2discrete4}. While the computation we carry out will be for the heat kernel over the spin--2 Laplacian, it will be sufficient for our purposes as the $E_6$ mode does not couple to the graviphoton flux. We will firstly compute the number of zero modes on $\ads2s2n$. To do so, we will use our usual definition of the number of zero modes, according to which
\begin{equation}
n_0=\sideset{}{'}\sum_{\ell}\int d^2x \sqrt{g} h_{\ell\,mn}\l(x\r)h_{\ell}^{*\,mn}\l(x\r)
\end{equation}
where the prime over the sum denotes that we will sum over only those $\ell$ modes which are integer multiples of $N$. These are the only modes which survive the orbifold projection. The integral is volume divergent, and as for the vector case, we regulate the divergence by cutting off the radial coordinate $\eta$ at a large value $\eta_0$. The regularized number of zero modes is the order 1 term in the large $\eta_0$ expansion for the above integral. 
We finally find that
\begin{equation}
n_0=\sideset{}{'} \sum_{\ell}\tanh^{2\ell}\l({\eta_0\over 2}\r)\l(2+4\ell{\cosh\eta_0\over\sinh^2\eta_0}+4\ell^2{1\over\sinh^2\eta_0}\r),
\end{equation}
Now we consider first the case where there is no orbifold imposed on the theory. In this case
\begin{equation}
n_0\simeq {3\over 2}e^{\eta_0}-3+\mathcal{O}\l(e^{-\eta_0}\r).
\end{equation}
Then the regulated number of zero modes is given by
\begin{equation}
n_0=-3,
\end{equation}
When we impose the $\zn$ orbifold, then the number of zero modes is given by
\begin{equation}
n_0=\sum_{p=1}^\infty\tanh^{2Np}\l({\eta_0\over 2}\r)\l(2+4Np{\cosh\eta_0\over\sinh^2\eta_0}+4N^2p^2{1\over\sinh^2\eta_0}\r),
\end{equation}
Now the sum over $p$ may be carried out, and the resultant expression be expanded in large $\eta_0$ to obtain
\begin{equation}
n_0= {3\over 2N}e^{\eta_0}-1+\mathcal{O}\l(\e^{-\eta_0}\r).
\end{equation}
Then the regularised number of zero modes (for any choice of $N\neq 1$) is 
\begin{equation}\label{nzeroe6}
n_0=-1.
\end{equation}
We will now compute the contribution of this discrete series to the heat kernel over non-zero modes on $\ads2s2n$. This requires us to pick out modes on $\ads2s2n$ which are invariant under the full $\zn$ orbifold. The analysis is identical to the vector mode analysis of \cite{Gupta:2013sva}, and we finally obtain
\begin{equation}
K= {1\over N}\sum_{s=0}^{N-1}\l(\sum_{\ell=2}^\infty\int^{\eta_0}\sqrt{g}d\eta d\theta\, g^{ac}g^{bd}\,h^{\ell}_{ab}h^{\ell\,*}_{cd}\l(e^{i\ell{2\pi s\over N}}+e^{-i\ell{2\pi s\over N}}\r)\r)\sum_{j=1}^\infty \chi_{j}\l({\pi s\over N}\r)e^{-{t\over a^2}j\l(j+1\r)}
\end{equation}
Now for the bracketed term, 
\begin{equation}
T\l(s\r)=\sum_{\ell=2}^\infty\int^{\eta_0}\sqrt{g}d\eta d\theta\, g^{ac}g^{bd}\,h^{\ell}_{ab}h^{\ell\,*}_{cd}\l(e^{i\ell{2\pi s\over N}}+e^{-i\ell{2\pi s\over N}}\r),
\end{equation}
when $s=0$, this is just given by the above counting of the number of zero modes in the unorbifolded theory.
\begin{equation}
T\l(0\r)={3\over 2}e^{\eta_0}-3+\mathcal{O}\l(e^{-\eta_0}\r).
\end{equation}
When $s$ is not equal to zero, we have
\begin{equation}\label{ts}
T\l(s\r)=-1-2\cos\l({2\pi s\over N}\r).
\end{equation}
Then the regularized form for the heat kernel is given by
\begin{equation}
K=-{3\over N}\sum_{j=1}^\infty\l(2j+1\r)e^{-{t\over a^2}j\l(j+1\r)}+{1\over N}\sum_{s=1}^{N-1}T\l(s\r)\sum_{j=1}^\infty\chi_{j}\l(\pi s\over N\r)e^{-{t\over a^2}j\l(j+1\r)},
\end{equation}
where we have discarded the volume divergence as per our usual procedure.
Using the expression \eqref{ts}, the heat kernel over the $j\geq 1$ modes may be written as
\begin{equation}
K=-{1\over N}\sum_{j=1}^\infty\sum_{s=0}^{N-1}\l(1+2\cos{2\pi s\over N}\r)\chi_{j}\l(\pi s\over N\r)e^{-{t\over a^2}j\l(j+1\r)}.
\end{equation}
Hence, the degeneracy of the eigenvalue $E_j=j\l(j+1\r)$ is given by
\begin{equation}\label{degE6l}
d\l(j\r)=-{1\over N}\sum_{s=0}^{N-1}\l(1+2\cos{2\pi s\over N}\r)\chi_{j}\l({\pi s\over N}\r),\quad j\geq 1,
\end{equation}
and
\begin{equation}\label{degE6l0}
d\l(j\r)=-1,\quad j=0,
\end{equation}
which is read off from the expression \eqref{nzeroe6} for the number of zero modes.
\section{Discrete Modes of the Gravitino}\label{discretesec}
In this section we will take into account the presence of discrete modes of the gravitino on $\adss$. We first review the origin of these modes. Just as we constructed a basis for the vector and the graviton by acting with derivatives on the scalar, we shall construct a basis for the spin-${3\over 2}$ field by acting with spinor covariant derivatives and gamma matrices on the spin-${1\over 2}$ eigenfunctions of the Dirac operator \cite{Banerjee:2011jp}. Consider first the spin-${1\over 2}$ eigenfunctions of the Dirac operator on $\adss$ \cite{Camporesi:1995fb}.
\begin{equation}\label{chi}
\begin{split}
\chi_{k}^{\pm}\l(\lambda\r) = {1\over \sqrt{4\pi a^2}}&\,
{\l\vert{\G\l(1+k+i\lambda\r)\over \G\l(k+1\r)\G\l({1\over 2}+i\lambda\r)}\r\vert}\,
e^{ i\left(k+{1\over 2}\right)\phi} \\&
\left(\begin{array}{cc} i{\lambda\over k+1}\, \sinh^{k+1}{\eta\over 2}\cosh^k {\eta\over 2}
F\l(k+1+i\lambda,k+1-i\lambda;k+2;-\sinh^2{\eta\over 2}\r) \\ \pm \cosh^{k+1}{\eta\over 2}\sinh^k{\eta\over 2} F\l(k+1+i\lambda,k+1-i\lambda;k+1;-\sinh^2{\eta\over 2}\r)
 \end{array}\right),
\end{split}
\end{equation}
and
\begin{equation}\label{eta}
\begin{split}
\eta_{l,m}^{\pm} = {1\over \sqrt{4\pi a^2}}&\,
{\l\vert{\G\l(1+k+i\lambda\r)\over \G\l(k+1\r)\G\l({1\over 2}+i\lambda\r)}\r\vert}\,
e^{ -i\left(k+{1\over 2}\right)\phi} \\&
\left(\begin{array}{cc} i\, \cosh^{k+1}{\eta\over 2}\sinh^k{\eta\over 2} F\l(k+1+i\lambda,k+1-i\lambda;k+1;-\sinh^2{\eta\over 2}\r)\\ \pm i{\lambda\over k+1}\cosh^k{\eta\over 2}\sinh^{k+1}{\eta\over 2}F\l(k+1+i\lambda,k+1-i\lambda;k+2;-\sinh^2{\eta\over 2}\r) \end{array}\right),
\end{split}
\end{equation}
where
\begin{equation}
k\in \mathbb{Z}, \quad 0\leq k<\infty, \quad 0<\lambda<\infty.
\end{equation}
These satisfy the eigenvalue equation
\begin{equation}
\slashed{D}_{\adss} \chi_{k}^\pm\l(\lambda\r) =\pm i\, a^{-1}\,\lambda \chi_{k}^\pm\l(\lambda\r)\, , \qquad
\not \hskip -4pt D_{\adss} \eta_{k}^\pm\l(\lambda\r) =\pm i\, a^{-1}\,
\lambda 
\eta_{k}^\pm\l(\lambda\r)\, .
\end{equation}
Then a basis for spin--$3/2$ fields on $\adss$ is given by
\begin{equation}\label{principal1}
\xi^{\pm}_m\l(\lambda\r)=D_m \chi^{\pm}, \quad \xi^{\pm}_m\l(\lambda\r)=\hat{\gamma}_m\chi^{\pm},
\end{equation}
and
\begin{equation}\label{principal2}
\hat{\xi}^{\pm}_m\l(\lambda\r)=D_m \eta^{\pm}, \quad \hat{\xi}^{\pm}_m\l(\lambda\r)=\hat{\gamma}_m\eta^{\pm}.
\end{equation}
Here $\hat{\gamma}_m$ are the two dimensional Dirac matrices, which we choose to be
\begin{equation}
\gamma_0=-\tau_2,\quad \gamma_1=\tau_1.
\end{equation}
The corresponding basis for four-dimensional Dirac matrices is given by
\begin{equation}
\gamma^0=-\sigma_3\otimes\tau_2,\quad\gamma^1=\sigma_3\otimes\tau_1,\quad\gamma^2=-\sigma_2\otimes\mathbb{I}_2,\quad \gamma^3=\sigma_1\otimes\mathbb{I}_2.
\end{equation}
We shall refer to a spin-${1\over 2}$ or a spin-${3\over 2}$ field expanded in a basis of \eqref{chi} and \eqref{eta} or \eqref{principal1} and \eqref{principal2} as belonging to the principal series. There is an additional set of normalizable modes on $\adss$ given by \cite{Camporesi:1995fb}
\begin{equation}\label{discreteads2}
\xi^{(k)\pm}_m=\l(D_m \pm {1\over 2a}\hat{\g}_m\r)\chi_k^{\pm}\l(i\r), \quad \hat{\xi}^{(k)\pm}_m=\l(D_m \pm {1\over 2a}\hat{\g}_m\r)\eta_k^{\pm}\l(i\r),
\end{equation}
which were not included in the basis \eqref{principal1}, \eqref{principal2} as $\chi\l(i\r)$ and $\eta\l(i\r)$ are not normalizable modes. These tensor with spin-${1\over 2}$ modes along $\ss2$ to furnish a set of modes for the spin-${3\over 2}$ field on $\adss\otimes\ss2$ given by 
\begin{equation}\label{discrete}
\xi^{(k)\pm}_n=\psi_{\ell m}\otimes\l(D_n \pm {1\over 2a}\sigma_3\g_n\r)\chi_k^{\pm}\l(i\r), \quad \hat{\xi}^{(k)\pm}_n=\psi_{\ell m}\otimes\l(D_n \pm {1\over 2a}\sigma_3\g_n\r)\eta_k^{\pm}\l(i\r).
\end{equation}
These modes furnish a basis for eigenmodes of the Dirac operator on $\adss\otimes\ss2$ with eigenvalues labeled by $\ell$. We shall refer to a wave function expanded in terms of the basis \eqref{discrete} as belonging to the discrete series. We now turn to computing the degeneracy of these eigenvalues. This will be useful for the computation of the one-loop determinant over these modes in the graviphoton background because though the eigenvalues get shifted by the graviphoton flux, the degeneracies do not change. Further, for computing the contribution of the fixed point terms to the log term, only the degeneracy is relevant. 
Now one may explicitly show that $\xi^{+}$ and $\xi^{-}$ are proportional to each other and $\hat{\xi}^{+}$ and $\hat{\xi}^{-}$ are proportional to each other. A convenient basis to use for computing the degeneracy is hence given by
\begin{equation}
\Psi_n= \psi_n^{(k)}\otimes\phi_{\ell m},\quad \vert k\vert=1,2,\ldots\infty,\, m\in \l[-\ell,\ell\r].
\end{equation}
If $k>0$ then $\psi_n$ is $\xi^{+}$ and if $k<0$ then $\psi_n$ is $\hat{\xi}^{+}$, and if $m>0$ then $\phi_{\ell m}$ is $\chi^{\pm}_{\ell m}$ and if $m<0$ then $\phi_{\ell m}$ is $\eta^{\pm}_{\ell m}$. To study the orbifold action, we define new variables $\tk=k+{1\over 2}$ and $\tm=m+{1\over 2}$. Now under the orbifold action $\l(\theta,\phi\r)\mapsto\l(\theta+{2\pi\over N},\phi-{2\pi\over N}\r)$ $\Psi_n$ transforms as
\begin{equation}
\Psi_n\mapsto e^{i{2\pi\l(\tk-\tm\r)\over N}}\Psi_n,
\end{equation}
hence the modes which survive the orbifold projection are those for which $\tk-\tm=Np$. Then the degeneracy of the eigenvalue labeled by $\ell$ is given by\footnote{We label eigenfunctions now by $\tk$ and $\tm$, rather than $t$ and $m$}
\begin{equation}
d_{\ell}=\sum_{\tk}\sum_{\tm}\l\langle\Psi_{\tk\tm}\vert \Psi_{\tk\tm}\r\rangle \delta_{\tk-\tm,Np},
\end{equation}
where the inner product is the Lorentz invariant dot product of the spinor with itself, integrated over the spacetime manifold. To avoid proliferation of indices, we have suppressed the vector index in $\Psi$. 
This may be evaluated to finally obtain
\begin{equation}
d_{\ell}={2\over N}\sum_{s=0}^{N-1}{\sin\l({2\pi s\over N}\l(\ell+1\r)\r)\over\sin\l({\pi s\over N}\r)} \sum_{k=1}^\infty\l(1+2k+\cosh\eta_0\r){\sinh^{2k}{\eta_0\over 2}\over\cosh^{2k+2}{\eta_0\over 2}}\cos\l({{2\pi s\over N}\l(k+{1\over 2}\r)}\r)
\end{equation}
where we have done the volume integral by cutting off the $\adss$ radius at a large value $\eta_0$, and have rewritten the sum over $\tk$ as a sum over $k$. Now the sum over $k$ may be carried out, and the expansion about large $\eta_0$ can be done to extract the order 1 term. 
We therefore find the degeneracy
\begin{equation}
d_{\ell}=-{4\over N}\sum_{s=0}^{N-1}{\sin\l({2\pi s\over N}\l(\ell+1\r)\r)\over\sin\l({\pi s\over N}\r)}\cos\l({\pi s\over N}\r)
\end{equation}
for the eigenvalue labeled by $\ell$. 
This may further be written as
\begin{equation}\label{fzndeg}
d_{\ell}={1\over N}d_{\ell,old}-{4\over N}\sum_{s=1}^{N-1}{\sin\l({2\pi s\over N}\l(\ell+1\r)\r)\over\sin\l({\pi s\over N}\r)}\cos\l({\pi s\over N}\r).
\end{equation}
The $\ell=0$ set of modes require special care. In particular, one may show, on following the above steps, that in the unorbifolded case the degeneracy is $-8$, whereas once the orbifold is imposed, it shifts to $-4$.
This will be taken into account when we compute the heat kernel for the graviphoton background.
\section{Zero Modes of the Gravitino in the Graviphoton Background}\label{fermzeromod}
The kinetic operator for the graviphoton background when evaluated over the above discrete modes also contains zero modes. We now explicitly evaluate the number of zero modes in this background using the results of \cite{Banerjee:2011jp}. Following \cite{Banerjee:2011jp} we consider a 10-dimensional spinor
\begin{equation}
\Psi_n= \psi_{\ell m}\otimes\psi_n\otimes\phi,
\end{equation}
where $\phi$ is a suitably chosen $SO(6)$ Dirac spinor which is an eigenfunction of $\Gamma^{45}$ with eigenvalue $i$, and $\psi_{\ell m}$ is a Dirac spinor on $\ss2$, and $\psi_n$ is the discrete mode along $\adss$. Then there are 4 linearly independent normalized zero modes of the graviphoton kinetic operator
\begin{equation}
\begin{split}
\Psi_m^{(A)}&={1\over 2}\l(-i+i\sigma_3-\hat{\G}^4-\sigma_3\hat{\G}^4\r)\chi^+_{00}\otimes\xi^{(k)+}_m\otimes\phi,\\
\Psi_m^{(B)}&={1\over 2}\l(-i+i\sigma_3-\hat{\G}^4-\sigma_3\hat{\G}^4\r)\eta^+_{00}\otimes\xi^{(k)+}_m\otimes\phi,\\
\Psi_m^{(C)}&={1\over 2}\l(i+i\sigma_3-\hat{\G}^4+\sigma_3\hat{\G}^4\r)\chi^+_{00}\otimes\hat{\xi}^{(k)+}_m\otimes\phi,\\
\Psi_m^{(D)}&={1\over 2}\l(i+i\sigma_3-\hat{\G}^4+\sigma_3\hat{\G}^4\r)\eta^+_{00}\otimes\hat{\xi}^{(k)+}_m\otimes\phi.\\
\end{split}
\end{equation}
We may further show that the inner products\footnote{These are the same inner products as defined in the previous section, they are the modulus squared of the relevant spinor, integrated over the spacetime manifold.} of these spinors with themselves are given by
\begin{equation}
\l\langle\Psi_m^{(A)}\vert\Psi_m^{(A)}\r\rangle= \l\langle\Psi_m^{(C)}\vert\Psi_m^{(C)}\r\rangle = \l\langle\xi^{(k)+}\vert\xi^{(k)+}\r\rangle,
\end{equation}
and
\begin{equation}
\l\langle\Psi_m^{(B)}\vert\Psi_m^{(B)}\r\rangle= \l\langle\Psi_m^{(D)}\vert\Psi_m^{(D)}\r\rangle = \l\langle\hat{\xi}^{(k)+}\vert\hat{\xi}^{(k)+}\r\rangle,
\end{equation}
where we have used the fact that the spinor components along the $\ss2$ and internal directions have inner product 1. 
We first calculate the number of zero modes in the unorbifolded theory and then impose the $\zn$ orbifold. In the unorbifolded theory, the number of zero modes is given by
\begin{equation}
n_0= \l\langle\Psi_m^{(A)}\vert\Psi_m^{(A)}\r\rangle+\l\langle\Psi_m^{(B)}\vert\Psi_m^{(B)}\r\rangle+ \l\langle\Psi_m^{(C)}\vert\Psi_m^{(C)}\r\rangle+\l\langle\Psi_m^{(D)}\vert\Psi_m^{(D)}\r\rangle,
\end{equation}
which is further given by
\begin{equation}
n_0=2\sum_{k=1}^\infty\l(\l\langle\xi^{(k)+}\vert\xi^{(k)+}\r\rangle+\l\langle\hat{\xi}^{(k)+}\vert\hat{\xi}^{(k)+}\r\rangle\r).
\end{equation}
These expressions are divergent due to the infinite volume of $\adss$, and will be regulated as per our usual prescription of putting a cutoff at a large $\adss$ radius $\eta_0$. We then find that the $\xi$ modes contribute to the number of zero modes via
\begin{equation}
n_0^{\xi}={1\over 2}\sum_{k=1}^\infty k\l(k+1\r)\int_0^{\eta_0}d\eta\,\sinh\eta {\sinh^{2k-2}{\eta\over 2}\over\cosh^{2k+4}{\eta\over 2}}={1\over 2}\sum_{k=1}^\infty\l(1+2k+\cosh\eta_0\r){\sinh^{2k}{\eta_0\over 2}\over\cosh^{2k+2}{\eta_0\over 2}},
\end{equation}
and the $\hat{\xi}$ modes contribute equally.
\begin{equation}
n_0^{\hat{\xi}}={1\over 2}\sum_{k=1}^\infty\l(1+2k+\cosh\eta_0\r){\sinh^{2k}{\eta_0\over 2}\over\cosh^{2k+2}{\eta_0\over 2}}.
\end{equation}
Doing the sum over $k$, we find that
\begin{equation}
n_0^{\xi}=n_0^{\hat{\xi}}=-1+\cosh\eta_0.
\end{equation}
The regularised number of zero modes is taken to be the order 1 term in the large $\eta_0$ expansion, so
\begin{equation}
n_0^{\xi}=n_0^{\hat{\xi}}=-1.
\end{equation}
We then find that the number of zero modes for the graviphoton background in $\adss\otimes\ss2$ is given by
\begin{equation}
n_0=-4.
\end{equation}
Now we impose the $\zn$ orbifold. The spinors on $\adss\otimes\ss2$ that correspond to zero modes of the graviphoton background are moded as follows
\begin{equation}
\begin{split}
\chi^+_{00}\otimes\xi^{(k)+}_m= e^{i{\phi\over 2}}e^{i\l({k+{1\over 2}}\r)\theta},&\quad \eta^+_{00}\otimes\hat{\xi}^{(k)+}_m= e^{-i{\phi\over 2}}e^{-i\l({k+{1\over 2}}\r)\theta},\\ \eta^+_{00}\otimes\xi^{(k)+}_m= e^{-i{\phi\over 2}}e^{i\l({k+{1\over 2}}\r)\theta},&\quad \chi^+_{00}\otimes\hat{\xi}^{(k)+}_m= e^{i{\phi\over 2}}e^{-i\l({k+{1\over 2}}\r)\theta}.
\end{split}
\end{equation}
Then it is easy to see that the modes which survive the projection $\l(\theta,\phi\r)\mapsto\l(\theta+{2\pi\over N},\phi-{2\pi\over N}\r)$ are given by $k=Np$ for the first line and $k=Np-1$ for the second line, where $p=1,2,\ldots$. The number of zero modes on the orbifolded geometry is then given by
\begin{equation}
n_0=\sum_{p=1}^\infty \l(\l\langle\xi^{(Np)+}\vert\xi^{(Np)+}\r\rangle+\l\langle\xi^{(Np-1)+}\vert\xi^{(Np-1)+}\r\rangle +\l\langle\hat{\xi}^{(Np)+}\vert\hat{\xi}^{(Np)+}\r\rangle+\l\langle\hat{\xi}^{(Np-1)+}\vert\hat{\xi}^{(Np-1)+} \r\rangle\r).
\end{equation}
Carrying out the sum over $p$ and retaining the order-1 term in the $\eta_0$ expansion, we find
\begin{equation}
n_0=-2.
\end{equation}
Thus the number of fermion zero modes is $-2$, which is half the number on the unorbifolded space \cite{Banerjee:2011jp}. This is consistent with the fact that the orbifold breaks half the supersymmetry of the $\adss\otimes\ss2$ graviphoton background.

\section{Useful Summation Formulae}\label{usefulsums}
In this appendix we list various formulae useful when summing over the conical terms. In the bosonic computation the following formulae will be useful,
\ben
&&\sum_{l=0}^\infty\chi_{l}\l(\frac{\pi s}{N}\r)=\frac{1}{2\sin^2\l[\frac{\pi s}{N}\r]},\qquad \sum_{l=1}^\infty\chi_{l}\l(\frac{\pi s}{N}\r)=\frac{\cos\l[\frac{2\pi s}{N}\r]}{2\sin^2\l[\frac{\pi s}{N}\r]},\nn\\
&&\sum_{l=0}^\infty\int_0^\infty d\lambda\chi^b_\lambda\l(\frac{\pi s}{N}\r)\chi_{l}\l(\frac{\pi s}{N}\r)=\frac{1}{4\sin^4\l[\frac{\pi s}{N}\r]},\nn\\
&&\sum_{l=1}^\infty\int_0^\infty d\lambda\chi^b_\lambda\l(\frac{\pi s}{N}\r)\chi_{l}\l(\frac{\pi s}{N}\r)=\frac{\cos\l[\frac{2\pi s}{N}\r]}{4\sin^4\l[\frac{\pi s}{N}\r]}.
\een
In the fermonic computation, the following formulae will be useful
\ben
&&\sum_{l=0}^\infty\chi_{l+\frac{1}{2}}\l(\frac{\pi s}{N}\r)=\frac{\cos\l[\frac{\pi s}{N}\r]}{2\sin^2\l[\frac{\pi s}{N}\r]},\qquad \sum_{l=1}^\infty\chi_{l+\frac{1}{2}}\l(\frac{\pi s}{N}\r)=\frac{\cos\l[\frac{3\pi s}{N}\r]}{2\sin^2\l[\frac{\pi s}{N}\r]},\nn\\
&&\sum_{l=0}^\infty\int_0^\infty d\lambda\chi^f_\lambda\l(\frac{\pi s}{N}\r)\chi_{l+\frac{1}{2}}\l(\frac{\pi s}{N}\r)=\frac{\cos^2\l[\frac{\pi s}{N}\r]}{4\sin^4\l[\frac{\pi s}{N}\r]},\nn\\
&&\sum_{l=1}^\infty\int_0^\infty d\lambda\chi^f_\lambda\l(\frac{\pi s}{N}\r)\chi_{l+\frac{1}{2}}\l(\frac{\pi s}{N}\r)=\frac{\cos\l[\frac{\pi s}{N}\r]\cos\l[\frac{3\pi s}{N}\r]}{4\sin^4\l[\frac{\pi s}{N}\r]}.
\een
In the above we have used,
\ben
&&\chi_{l}\l(\frac{\pi s}{N}\r)=\frac{\sin\l[\frac{\pi s(1+2l)}{N}\r]}{\sin\l[\frac{\pi s}{N}\r]},\qquad \chi^b_\lambda\l(\frac{\pi s}{N}\r)=\frac{\cosh\l(\pi-\frac{2\pi s}{N}\r)\lambda}{\cosh(\pi\lambda)\sin\l(\frac{\pi s}{N}\r)},\nn\\
&&\chi_{l+\frac{1}{2}}\l(\frac{\pi s}{N}\r)=\frac{\sin\l[\frac{2\pi s(1+l)}{N}\r]}{\sin\l[\frac{\pi s}{N}\r]},\qquad \chi^f_\lambda\l(\frac{\pi s}{N}\r)=\frac{\sinh\l(\pi-\frac{2\pi s}{N}\r)\lambda}{\sinh(\pi\lambda)\sin\l(\frac{\pi s}{N}\r)}.
\een
We will also need the following summation formulae
\ben
&&\sum_{s=1}^{N-1}\frac{1}{\sin^2\l[\frac{\pi s}{N}\r]}=\frac{N^2-1}{3},\qquad \sum_{s=1}^{N-1}\frac{1}{\sin^4\l[\frac{\pi s}{N}\r]}=\frac{N^4+10N^2-11}{45},\nn\\
&&\sum_{s=1}^{N-1}\frac{\cos\l(\frac{2\pi s}{N}\r)}{\sin^2\l(\frac{\pi s}{N}\r)}=\frac{N^2-6N+5}{3},\qquad \sum_{s=1}^{N-1}\frac{\cos\l(\frac{4\pi s}{N}\r)}{\sin^2\l(\frac{\pi s}{N}\r)}=\frac{N^2-12N+23}{3},\nn\\
&&\sum_{s=1}^{N-1}\frac{\cos\l(\frac{2\pi s}{N}\r)}{\sin^4\l(\frac{\pi s}{N}\r)}=\frac{19 - 20 N^2 + N^4}{45},\qquad \sum_{s=1}^{N-1}\frac{\cos\l(\frac{4\pi s}{N}\r)}{\sin^4\l(\frac{\pi s}{N}\r)}=\frac{N^4-110N^2+360N-251}{45},\nn\\
&&\sum_{s=1}^{N-1}\sin^2\l[\frac{\pi s}{N}\r]=\frac{N}{2},\qquad \sum_{s=0}^{N-1}\cos^2\l[\frac{\pi s}{N}\r]=\frac{N}{2}.
\een
\bibliography{paper}
\bibliographystyle{unsrt}
\end{document}